\documentclass[titlepage,10pt,leqno]{article}
\usepackage{xcolor}
\usepackage{latexsym}
\usepackage{amsfonts}
\usepackage{amsmath}
\usepackage{amssymb}
\usepackage{pifont}

\newlength{\filength}
\newsavebox{\gcbox}
\sbox{\gcbox}{\framebox[\filength]{\rule{0ex}{2ex}}}

  \newtheorem{theorem}{Theorem}[section]
  \newtheorem{corollary}[theorem]{Corollary}
  \newtheorem{claim}[theorem]{Claim}

\newcommand\qedblob{\ding{113}}
\def\literalqed{{\ \nolinebreak\hfill\mbox{\qedblob\quad}}}

  \newtheorem{lemma}[theorem]{Lemma}

  \newtheorem{definition}[theorem]{Definition}
  \newtheorem{example}[theorem]{Example}
  \newtheorem{construction}[theorem]{Construction}
\newcount\hour  \newcount\minutes  \hour=\time  \divide\hour by 60
\minutes=\hour  \multiply\minutes by -60  \advance\minutes by \time
\def\mmmddyyyy{\ifcase\month\or Jan\or Feb\or Mar\or Apr\or May\or Jun\or Jul\or
  Aug\or Sep\or Oct\or Nov\or Dec\fi \space\number\day, \number\year}
\def\hhmm{\ifnum\hour<10 0\fi\number\hour :%
  \ifnum\minutes<10 0\fi\number\minutes}

\makeatletter
\def\@citex[#1]#2{\if@filesw\immediate\write\@auxout{\string\citation{#2}}\fi
  \def\@citea{}\@cite{\@for\@citeb:=#2\do
    {\@citea\def\@citea{,\linebreak[0]}\@ifundefined
       {b@\@citeb}{{\bf ?}\@warning
       {Citation `\@citeb' on page \thepage \space undefined}}%
\hbox{\csname b@\@citeb\endcsname}}}{#1}}
\newcommand{\singlespacing}{\let\CS=
\@currsize\renewcommand{\baselinestretch}{1}\tiny\CS}
\newcommand{\singlespacingplus}{\let\CS=
\@currsize\renewcommand{\baselinestretch}{1.25}\tiny\CS}
\newcommand{\doublespacing}{\let\CS=
\@currsize\renewcommand{\baselinestretch}{1.75}\tiny\CS}
\newcommand{\extradoublespacing}{\let\CS=
\@currsize\renewcommand{\baselinestretch}{1.9}\tiny\CS}
\newcommand{\draftspacing}{\let\CS=
\@currsize\renewcommand{\baselinestretch}{2.0}\tiny\CS}
\newcommand{\hugedraftspacing}{\let\CS=
\@currsize\renewcommand{\baselinestretch}{2.4}\tiny\CS}
\newcommand{\normalspacing}{\singlespacing}
\makeatother%

\normalspacing



\newcommand{\p}{{\rm P}}

\newcommand{\np}{{\rm NP}}

\newenvironment{proofs}{\noindent{\bf Proof.}\hspace*{1em}}{\literalqed\bigskip}

\newcommand{\condition}{\,\mid \:}
 
\def\land{{\; \wedge \;}}
\def\lor{{\; \vee \;}}

\newcommand{\seq}{\subseteq}

\def\parpair#1{{{(\!\!~#1~\!\!)}}}

\newcommand{\nominee}[1]{{{\mbox{\it{Nominees}}(#1)}}}

\newcommand{\score}[1]{{{\mbox{\it{score}}(#1)}}}
\newcommand{\scoresub}[2]{{{\mbox{\it{score}}_{#1}(#2)}}}
\newcommand{\diff}[1]{{{\mbox{\it{diff}}(#1)}}}
\newcommand{\surplus}[1]{{{\mbox{\it{surplus}}(#1)}}}

\title{Anyone but Him: The Complexity of Precluding an Alternative\thanks{Supported 
in part by
 grants NSF-CCR-0311021,
   NSF-CCF-0426761, and DFG-RO-1202/9-1.
 A preliminary version of this paper appeared in 
AAAI-05~\cite{hem-hem-rot:c:destructive-control}.
 This work was done in part while the
 authors were visiting 
   Julius-Maximilians-Universit\"at~W\"{u}rzburg,
 and while the first author was on sabbatical at the
 University of Rochester.}}
\author{Edith Hemaspaandra\thanks{URL: \mbox{\tt{}www.cs.rit.edu/$\tilde{~}$eh}.} \\
Department of Computer Science \\
Rochester Institute of Technology \\
Rochester, NY 14623, USA
\and
Lane A. Hemaspaandra\thanks{URL: \mbox{\tt{}www.cs.rochester.edu/u/lane}.}
\\Department of Computer Science \\
University of Rochester \\
Rochester, NY 14627, USA
\and
J\"{o}rg Rothe\thanks{URL: \mbox{\tt{}wwwold.cs.uni-duesseldorf.de/$\tilde{~}$rothe}.} \\
Institut f\"{u}r Informatik \\
Heinrich-Heine-Universit\"{a}t D\"{u}sseldorf \\
40225 D\"{u}sseldorf, Germany
}

\date{July 28, 2005; revised March~4, 2026}

\begin{document}

\sloppy

\maketitle

\begin{abstract}
Preference aggregation in a multiagent setting is a central issue in
both human and computer contexts.  In this paper, we study in terms of
complexity the vulnerability of preference aggregation to destructive
control.  That is, we study the ability of an election's chair to,
through such mechanisms as voter/candidate
addition/suppression/partition, ensure that a particular candidate
(equivalently, alternative) does not win.  And we study the extent to
which election systems can make it impossible, or computationally
costly (NP-complete), for the chair to execute such control.  Among the
systems we study---plurality, Condorcet, and approval voting---we find
cases where systems immune or computationally resistant to a
chair choosing the winner nonetheless are vulnerable to the chair
blocking a victory.
Beyond that, we see that among our studied
systems
no one
system offers the best protection
against
destructive control.
Rather, the choice of a preference 
aggregation system will depend closely on which types of control one
wishes to be protected against.
We also find concrete cases where the 
complexity of or susceptibility to control varies dramatically 
based on the choice 
among natural tie-handling rules.
\medskip
\\[2mm]

\bigskip
\noindent\textcolor{red}{Note: This revision---the March 2026 Version 5---is identical to the
March 2006
Version~4 except in providing, as Appendix~A, a correction
to the second half of the proof of Theorem~4.21 as it appears
in both Version~4 and the AIJ~journal version; this proof also
replaces the analogous
proof part of
Theorem~6 of the AAAI version.}

\end{abstract}

\section{Introduction}

Voting systems 
provide 
a broad model for aggregating preferences in a multiagent
setting.  
The literature on 
voting is vast 
and 
active, and 
spans such 
areas as AI, complexity, economics, 
operations research, and political science.
As noted by Conitzer, Lang, and
Sandholm~\cite{con-lan-san:c:how-many-manipulate},
voting has been proposed as a mechanism for use in decision-making in
various computational settings, including
planning~\cite{eph-ros:c:clarke-tax,eph-ros:c:multiagent-planning} and
collaborative filtering~\cite{gil-hor-pen:c:collaborative-filtering}.
Voting also may be useful in many large-scale computer settings.
Examples of much recent interest include the (web-page) rank aggregation 
problem, and related issues of reducing ``spam'' results
in web search and improving similarity search,
for which the use of voting systems
has 
been proposed~\cite{dwo-kum-nao-siv:c:rank-aggregation,fag-kum-siv:c:similarity-search}.
In
such an automated 
setting, it is natural to imagine decisions with thousands or
millions of ``voters'' and ``candidates.''

In 
Bartholdi, Tovey, and Trick's 
seminal paper ``How hard is it to control an
election?''~\cite{bar-tov-tri:j:control}, 
the issue of constructive
control of election systems is studied: 
How hard is it for a chair (who 
knows all voters' preferences) to---through
control of
the voter or candidate set or of the partition structure of an
election---cause a given candidate (equivalently, alternative) to be the
(unique) winner?\footnote{In their model, which 
is also adopted here, the chair has complete
information on the voters' preferences.
This is a natural assumption
in many situations.  For example, in a computer science department,
after endless discussions, most
 people know what each person's position is on
key issues.  Also, 
since the case where
complete information is available to the chair is a special subcase of
the more general setting that allows information to
be specified with any level of completeness,
lower bounds obtained in the complete information setting
are inherited by any natural incomplete information model.}
Bartholdi, Tovey, and Trick
studied plurality and Condorcet voting, and seven natural types of
control: adding candidates, suppressing candidates, partition of candidates,
run-off partition of candidates, adding voters, suppressing voters, and
partition of voters.  They found that in some cases there is 
\emph{immunity} to
constructive control (if his/her candidate was not already 
the\footnote{Really ``\emph{a} unique winner,'' since there 
may 
be no winner at all, but we'll usually write ``\emph{the} 
unique winner'' when this is clear from context.}
unique winner,
no action of the specified type by the chair can make the candidate the unique
winner), in some cases there is 
\emph{(computational) resistance} to constructive
control (it is $\np$-complete to decide whether the chair can achieve his/her desired
outcome), and in some cases the system is 
\emph{(computationally) vulnerable} to
constructive control (there is a polynomial-time algorithm that will tell the
chair how to 
achieve the desired outcome whenever possible\footnote{This is
  more like
``computationally
  certifiably-vulnerable,'' see
  Definition~\ref{def:immune-resistant-vulnerable}.  Vulnerability as defined
  in~\cite{bar-tov-tri:j:control} 
  means one can quickly decide if there
  {\em exists\/} a way for the chair to achieve the desired outcome.%
}).

In this paper, we obtain results for each of 
their 14 cases (two preference
aggregation systems, each under seven control schemes) in the setting
of {\em destructive\/} control.  In contrast with constructive
control, in which a chair tries to ensure that a specified desirable
candidate is the (unique) winner, in destructive control the chair tries
to ensure that a specified detested candidate is not the (unique)
winner.  Regarding the naturalness of destructivity, the 
light-hearted
title of this paper tries to reflect the fact that, in human terms, one often
hears feelings expressed that focus strategically on precluding one candidate,
and of course in other settings this also may be a goal.  Regarding the
reality of electoral control, from targeted ``get-out-the-vote''
advertisements of parties and candidates to (alleged) voter suppression
efforts by independent groups, from the way a committee chair groups
alternatives to any case where a faculty member hands out student course
evaluations on a day some malcontent students are not in class, it is hard to
doubt that the desire for electoral control---both destructive and
constructive---is a real one.

Destruction has been previously studied by Conitzer, Lang, and
Sandholm~\cite{con-san:c:few-candidates,con-lan-san:c:how-many-manipulate}, 
but 
in the setting of election \emph{manipulation}---in which some
(coalition of)
voters knowing all other voters' preferences are free to shift their
own preferences to affect the outcome.  In contrast, in this
paper we study destruction in the very different
setting of electoral \emph{control}~\cite{bar-tov-tri:j:control}---where 
a chair, given
fixed and unchangeable voter preferences, tries to influence the outcome via
procedural/access means.

\begin{table*}[t]
\footnotesize
\centering
\begin{tabular}{||l||l|l||l|l||l|l||}
\hline\hline
                    & \multicolumn{2}{c||}{Plurality}
                    & \multicolumn{2}{c||}{Condorcet}
                    & \multicolumn{2}{c||}{Approval} \\ \hline
Control by          & Construct. & Destruct.
                    & Construct. & Destruct.
                    & Construct. & Destruct.    \\ \hline\hline
Adding Candidates   & R                     & {\bf R}    
                    & I                     & {\bf V}
                    & {\bf I}                 & {\bf V}
\\ \hline
Deleting Candidates & R                     & {\bf R}
                    & V                     & {\bf I}
                    & {\bf V}
& {\bf I}             \\ \hline
Partition           & TE: R                 & {\bf TE:} {\bf R}
                    & V                     & {\bf I}
                    & {\bf TE:} {\bf V}
& {\bf TE:} {\bf I}    \\
of Candidates       & TP: R                 & {\bf TP:} {\bf R} 
                    &                       &
                    & {\bf TP:} {\bf I}             & {\bf TP:} {\bf I}    \\ \hline
Run-off Partition   & TE: R                 & {\bf TE:} {\bf R} 
                    & V                     & {\bf I}
                    & {\bf TE:} {\bf V}
& {\bf TE:} {\bf I}   \\ 
of Candidates       & TP: R                 & {\bf TP:} {\bf R} 
                    &                       & 
                    & {\bf TP:} {\bf I}             & {\bf TP:} {\bf I}    \\ \hline
Adding Voters       & V                     & {\bf V}
                    & R                     & {\bf V}
                    & {\bf R}                 & {\bf V}
\\ \hline
Deleting Voters     & V                     & {\bf V}
                    & R                     & {\bf V}
                    & {\bf R}                 & {\bf V}
\\ \hline
Partition           & {\bf TE:} {\bf V}
& {\bf TE:} {\bf V}
                    & R                     & {\bf V}
                    & {\bf TE:} {\bf R}             & {\bf TE:} {\bf V}
\\
of Voters           & {\bf TP:} {\bf R}             & {\bf TP:} {\bf R}
                    &                       &
                    & {\bf TP:} {\bf R}             & {\bf TP:} {\bf V}
\\ 
\hline\hline
\end{tabular}
\caption{Summary of results.
Results new
to this paper are in boldface.  Nonboldface results
are due to Bartholdi, Tovey, and 
Trick~\cite{bar-tov-tri:j:control}.
Key: I = immune, R = resistant,
V = vulnerable, 
TE = Ties-Eliminate, TP = Ties-Promote.  
\label{tab:results}
}
\end{table*}

One might ask, ``Why bother studying destructive control, since any
rational chair would prefer to assert constructive control?''  The
answer is that it is plausible---and our results show it is indeed the
case---that destructive control may be possible in settings in which
constructive control is not.  Informally put, destructive control may
be easier for the chair to assert.  For example, we prove formally
that of the seven types of constructive control of Condorcet
elections that Bartholdi, Tovey, 
and Trick~\cite{bar-tov-tri:j:control} study, 
the four they showed not vulnerable to constructive control
are all vulnerable to destructive 
control.  The
remaining three cases regarding Condorcet voting
are vulnerable to constructive 
control~\cite{bar-tov-tri:j:control}, but we show that they 
are immune to destructive
control.\footnote{Savvy readers may wonder
whether 
there is something very troubling in having a system
be vulnerable to constructive control but immune 
to destructive control.
After all, to ensure that
the despised candidate $c$ is not the unique winner we simply have to
ask whether 
either at least one of the other candidates can be ensured to
unique-win-or-tie-for-winner or it can be ensured that there
are no winners.
Put somewhat formally, this implies 
that for \emph{strongly voiced}
(i.e., systems for which
whenever there is at least one candidate there 
will be at least one winner) election systems---though of 
course Condorcet voting is not strongly voiced and so this 
is not an issue for the three cases mentioned in the 
main text---destructive 
control polynomial-time disjunctively
truth-table reduces~\cite{lad-lyn-sel:j:com} to constructive control 
(redefined to speak not of ``unique winner'' but to speak
of ``winner (possibly with others also winning)''),
and so the destructive control problem can (within 
a polynomial factor) be no harder
\emph{computationally} than the (redefined)
constructive control problem (this reduction is 
noted in a different setting by
Conitzer and Sandholm~\cite{con-san:c:few-candidates}).
Our brief explanation of why cases of such a form would
not cause a paradox lies in the 
word ``computational'':  Although immunity is the most desirable
case in terms of 
security from control, 
the \emph{complexity} of recognizing whether a given candidate
can be precluded from winning in immune cases will most typically 
be in P---after all, we can \emph{never}, when immunity holds, 
change a given candidate from unique winner to not the 
unique winner, so the related 
decision problem is typically easy.  
(Technical side 
remark: We say ``will most typically be in P/is typically'' rather than
``will be in P/is'' 
because 
for
impractical
systems that---unlike those here---have winner-testing
problems that are not in P, it is 
in concept possible that one can have 
immunity and yet also have the related language problem not belong to P.)

The disjunctive-truth-table connection mentioned above explains why,
if $\p \neq \np$, it is impossible 
for any strongly voiced election system 
to have computational resistance to
destructive control hold for any problem that, when redefined to
embrace ties, is vulnerable to constructive control.\label{foo:longfootnote}%
}

Table~\ref{tab:results} summarizes our results on the
complexity of  destructively
controlling 
Condorcet, plurality, and approval elections.  We also
when needed 
obtain, for comparative purposes, new results on the complexity 
of constructive 
control, and Table~\ref{tab:results} displays those and also
constructive control results of
Bartholdi, Tovey, and Trick~\cite{bar-tov-tri:j:control}.
All
entries in boldface in Table~\ref{tab:results}
are new results obtained in this paper; the 
other results are due to
Bartholdi, Tovey, and Trick~\cite{bar-tov-tri:j:control}.
For each boldface ``V'' in the table, ``certifiably-vulnerable''
is in fact also achieved by our theorems.  We mention in 
passing that 
for nonboldface ``V''s in the table, ``certifiably-vulnerable''
can be seen directly from or by modifying the 
algorithms
of Bartholdi, Tovey, and Trick~\cite{bar-tov-tri:j:control}.

For control-by-partition problems---which will involve
subelection(s)---we distinguish between the models Ties-Eliminate
(TE, for short) and Ties-Promote (TP, for short), which 
define what happens when there are ties among winners 
in a subelection (before 
the final election), namely, all participating candidates 
are eliminated (TE), or all who tie for winner move forward (TP).
Note that these models do not apply to Condorcet voting, 
under which when a winner exists s/he is inherently 
unique; so the TE/TP 
distinction is made only for plurality and approval voting.  

The natural conclusion to draw from our results
is that when selecting an election/preference
aggregation system, one should at least be aware of the issue of the system's
vulnerability to control---and, beyond that, one's choice of system will
depend closely on which types of immunity or computational resistance one most
values.  Our results also show that constructive and destructive 
control often differ greatly:
A system immune
to constructive control may be vulnerable to destructive control,
and vice versa.  
Finally, our results show---in contrast with 
some comments in earlier papers---that breaking ties is 
far from a minor issue: For both voting types where tie-handling
rules are meaningful, we find cases where the 
complexity of or susceptibility to control varies dramatically 
based on the choice among natural tie-handling rules.

\section{Preliminaries%
}

We first define the three voting systems considered.  In \emph{approval
  voting}, each voter votes ``Yes'' or ``No'' for each candidate.
(So, for approval voting, a voter's preferences are reflected by a 0-1
vector.)  All candidates with the maximum number of ``Yes'' votes are
winners.
Approval voting has been proposed as a variant of plurality voting,
see Brams and Fishburn~\cite{bra-fis:b:approval-voting}.

Plurality and Condorcet voting are defined in terms 
of strict preferences.  For them, 
an election is given by a {\em preference profile}, a pair
$(C,V)$ such that $C$ is a set of candidates and $V$ is the
multiset (henceforth, we'll just say set, as a shorthand) 
of the voters' preference orders on~$C$.\footnote{
In various settings involving subelections, adding candidates, and resistance
constructions, we will speak of an election  $(C',V)$ where the preferences
of $V$ are over some $C \supseteq C'$.  In such cases, we intend the natural
interpretation: For the purpose of that election one views the
induced preference
order (or approval vector) for the restriction to $C'$.}
We assume that the
preference orders are irreflexive and antisymmetric (i.e., every voter
has strict preferences over the candidates), complete (i.e., every
voter ranks each candidate), and transitive.  

A {\em voting system\/}
is a rule for how to determine the winner(s) of an election.
Formally, any voting system is defined to be a {\em (social choice)
function\/} mapping any given preference profile (or the analog
with voters' 0-1 vectors for the approval voting case) to society's
aggregate {\em choice set}, the set of candidates who have won the
election.  

In {\em plurality voting}, each candidate with a maximum number of
``first preference among the candidates in the election'' voters for 
him/her wins.
In \emph{Condorcet voting}, 
for each $c \in C$, $c$ is a winner if and
only 
if for each $d \in C$ with $d \neq c$, 
$c$ 
defeats $d$ by a strict majority of votes
in a pairwise
election between them based on the voters' preferences.

The Condorcet Paradox observes that whenever there are at least
three candidates, due to cyclic aggregate preference rankings
Condorcet winners may not exist~\cite{con:b:condorcet-paradox}.  That
is, the set of winners may be empty.  However, a Condorcet winner
is unique whenever one does exist.  In the case of plurality and
approval voting, due to ties, there may exist multiple winners.  
Regarding ties, we---following
Bartholdi, Tovey, 
and Trick~\cite{bar-tov-tri:j:control} to best allow
comparison---focus in our 
control problems on creating a \emph{unique} winner (constructive),
and precluding a candidate from being the \emph{unique} winner
(destructive).   (Ties in subelections, for the partition problems,
are handled via the TE and TP rules described earlier.)

\section{Results}
\label{sec:results}

The issue of control of an election by the authority conducting it (called
the chair) can be studied under a variety of models and scenarios.  For
plurality and Condorcet voting, 
Bartholdi, Tovey, and Trick~\cite{bar-tov-tri:j:control}---and
for the rest of this section that paper will be referred to as 
``BTT92''---study constructive control
by adding candidates, deleting candidates, partition of candidates, run-off
partition of candidates, adding voters, deleting voters, and partition of
voters.
In their setting, the chair's goal is to
make a given candidate uniquely win the election.  Analogously, we consider 
in turn the
corresponding 
seven {\em destructive\/} control problems, where the chair's goal is
to preclude a given candidate from being the unique winner.  
For each of these
control scenarios,
we define the problem and present prior results and our results.
(Formally, for each type of control one
defines a decision problem and studies its
computational complexity.)
To make comparisons as easy as possible, we in stating these 
control problems
whenever possible exactly follow BTT92's
wording for constructive control
(except modified to the destructive case for the destructive cases),
and when we diverge, we explain why and how.

\subsubsection*{Control by Adding Candidates}

As is common, we state our decision problems as ``Given'' instances,
and a related Yes/No question.  The language in each case is the set
of all instances for which the answer is Yes.  
Since in each control scenario, the
``Given'' instance is identical for the constructive and the destructive case,
we state it just once and then state the corresponding two questions, one
for constructive and one for destructive control.

\begin{description}
\item[Given:] A set $C$ of qualified candidates and a distinguished candidate
  $c \in C$, a set $D$ of possible spoiler candidates, and a 
  set $V$ of voters with preferences (in the approval
case, the ``preferences'' will, as always for that case, actually be 
0-1 vectors)
over $C \cup D$.
\item[Question (constructive):] Is there a choice of candidates from
  $D$ whose entry into the election would assure that $c$ is the
  unique winner?
\item[Question (destructive):] Is there a choice of candidates from
  $D$ whose entry into the election would assure that $c$ is not the
  unique winner?
\end{description}

The above type of control captures the idea that the chair
tries to enthrone the desired candidate $c$ (in the constructive case)
or to dethrone the despised candidate $c$ (in the destructive case) by
introducing new ``spoiler'' candidates.

$V$ is formally a multiset.  However, throughout this paper we assume---as
is the standard approach in papers on the computational complexity
of elections---that in the input the preferences are coded
as a list (the ballots), one voter at a time, and 
in particular are not encoded as a multiset that uses
binary numbers to code cardinalities.

With this first problem---Control by Adding Candidates---stated, 
now is a good time to define our 
notions of control.
Our
terminology will closely follow
the notions
in BTT92, to allow comparison.

\begin{definition}
\label{def:immune-resistant-vulnerable}
  We say that a voting system is {\em immune to 
    control\/} in a given model of control (e.g., ``destructive 
control via adding candidates'') if
the model regards constructive control and 
it is never possible for the chair to 
by using his/her allowed model of control
change a
given candidate from being not a unique winner to being the
unique winner, or the model regards 
destructive control 
and it is never possible for the chair to 
by using his/her allowed model of control
change a
given candidate from being the unique winner to not 
being a 
unique winner.
If a system is not 
immune to a type of control, it is said to be 
\emph{susceptible} to that type of control. 

  A voting system 
  is said to be {\em (computationally)
    vulnerable to control\/} if it is susceptible to control 
  and the corresponding language problem is computationally
  easy (i.e., solvable in polynomial time).  If a system is not just
  vulnerable regarding some particular model of 
  control but one can even
produce in polynomial time the actual
  action of the chair to execute control the ``best'' way (namely, by adding
  or deleting the smallest number of candidates or voters for add/delete
  problems; for partition problems, any legal partition that works is 
  acceptable), we say the system is
  {\em (computationally) certifiably-vulnerable to 
(that model of) control}.\footnote{For the
problems studied here, certifiably-vulnerable 
implies vulnerable (but we list both, since if one 
studied add/delete problems stated not in terms 
of ``is there some subset'' or ``by adding/deleting at most~$k$''
but rather in terms of ``by adding/deleting exactly~$k$,''
then for certain systems the implication need not hold).}  

  A voting system 
is said to be {\em resistant to
    control\/} if 
it is susceptible to control 
but the corresponding language 
problem is computationally hard (i.e.,
  $\np$-complete).\footnote{It would be more natural
to define resistance as meaning the corresponding 
language is (many-one) $\np$-hard.  
However, in this paper,
we define resistance in terms of $\np$-completeness.  One reason is that
this matches
the way the term is used by BTT92.  More importantly, 
all the problems discussed in this paper have obvious $\np$
upper bounds since testing whether a given candidate 
has won a given election for the systems 
considered here is obviously in $\p$.  So for the problems in this paper, 
$\np$-completeness and $\np$-hardness stand or fall 
together.  We mention in passing that there are natural
election systems whose complexity seems beyond 
$\np$.  The first such case established was 
for the election system defined by Lewis Carroll 
in 1876~\cite{dod:unpubMAYBE:dodgson-voting-system}, 
where even the complexity of determining whether 
a given candidate has won is now known to be hard for
parallel access to $\np$~\cite{hem-hem-rot:j:dodgson}.
Other election systems whose winner complexity is hard for 
parallel access to $\np$ include
Kemeny and Young elections, see~\cite{hem-hem:c:computational-politics,spa-vog:c:theta-two-classic,spa-vog:c:kemeny,rot-spa-vog:j:young,hem-spa-vog:j:kemeny}.}
\end{definition}

For 
general background on 
the theory of $\np$-completeness, see, e.g.,~\cite{gar-joh:b:int,hop-ull:b:automata}.  

As to what is known about Constructive Control by Adding Candidates, 
BTT92 shows that plurality is resistant and 
Condorcet is immune.  Our results are:

\begin{theorem}
\label{thm:adding-candidates}
Approval (voting) is immune to constructive control by 
adding candidates, and plurality, Condorcet, and 
approval (voting) are respectively resistant, 
vulnerable/certifiably-vulnerable, and 
vul\-ner\-a\-ble/certi\-fia\-bly-vul\-ner\-a\-ble to destructive control
by adding candidates.
\end{theorem}

So, though Condorcet and approval are immune to constructive control
of this sort, they both are vulnerable to destructive control.
This reverses itself for:

\subsubsection*{Control by Deleting Candidates}

\begin{description}
\item[Given:] A set $C$ of candidates, a distinguished candidate $c \in C$, a
  set $V$ of voters, and a positive integer $k < ||C||$.
\item[Question (constructive):] Is there a set of $k$ or fewer candidates in
  $C$ whose disqualification would assure that $c$ is the unique
  winner?
\item[Question (destructive):] Is there a set of $k$ or fewer candidates in
  $C - \{c\}$ whose disqualification would assure that $c$ is not
  the unique winner?
\end{description}

In this type of control, the chair seeks to influence the outcome of the
election by suppressing certain candidates (other than~$c$),
in hopes that their voters now support $c$ to ensure $c$'s victory (in the
constructive case) or that they now support another
candidate to ensure stopping $c$ (in the destructive case).  
Note that this formalization of the destructive case is not a
perfect analog of the constructive case
of BTT92 in that
we explicitly prevent deleting~$c$, since otherwise
any voting system in which the winners can efficiently be determined would 
be 
trivially vulnerable to this type of control.

Here, BTT92 establishes for constructive control resistance 
for plurality and vulnerability
for Condorcet.  Our results are:

\begin{theorem}
\label{thm:deleting-candidates}
Approval is 
vulnerable/certifiably-vulnerable
to constructive control by 
deleting candidates.  Plurality, Condorcet, and 
approval are respectively resistant, 
immune, and immune 
to destructive control
by deleting candidates.\footnote{For this and all other problems
whose statements invoke a ``$k$'' bound, by immune we mean
that for no election (and thus no $k$) can the chair's action
ever cause change of the sort required to break immunity
(i.e., 
taking someone who is not a unique winner and making 
him/her be a unique winner in the constructive cases, or
taking someone who is a unique winner and making 
him/her no longer be a unique winner in the destructive cases),
and by susceptible we mean ``not immune'' (under the definition just
given).
}
\end{theorem}

We now handle jointly the two types of partition of candidates,
since they yield identical results.

\subsubsection*{Control by Partition of Candidates}

\begin{description}
\item[Given:] A set $C$ of candidates, a distinguished candidate $c \in C$,
  and a set $V$ of voters.
\item[Question (constructive):] Is there a partition of $C$ into $C_1$
  and $C_2$ such that $c$ is the unique winner in the sequential
  two-stage election in which the winners in the subelection $(C_1,V)$
  who survive the tie-handling rule move forward to face the
  candidates in $C_2$ (with voter set~$V$)?
\item[Question (destructive):] Is there a partition of $C$ into $C_1$
  and $C_2$ such that $c$ is not the unique winner in the sequential
  two-stage election in which the winners in the subelection $(C_1,V)$
  who survive the tie-handling rule move forward to face the
  candidates in $C_2$ (with voter set $V$)?
\end{description}

\subsubsection*{Control by Run-Off Partition of Candidates}

\begin{description}
\item[Given:] A set $C$ of candidates, a distinguished candidate $c \in C$,
  and a set $V$ of voters.
\item[Question (constructive):] Is there a partition of $C$ into $C_1$
  and $C_2$ such that $c$ is the unique winner of the election in
  which those candidates surviving (with respect to the tie-handling
  rule) subelections $(C_1,V)$ and $(C_2,V)$ have a run-off with voter
  set~$V$.
\item[Question (destructive):] Is there a partition of $C$ into $C_1$
  and $C_2$ such that $c$ is not the unique winner of the election in
  which those candidates surviving (with respect to the tie-handling
  rule) subelections $(C_1,V)$ and $(C_2,V)$ have a run-off with voter
  set $V$.
\end{description}

These two types of control express 
settings---one via a cascading setup, and one 
via a run-off setup---in which the chair tries to, overall,
partition the candidates in such a clever way that the
favored candidate $c$ is made the unique winner (in the constructive case)
or that the hated candidate $c$
fails to be the unique winner (in the destructive case).
Here, BTT92 shows that 
for constructive control 
plurality is resistant (and their result on that holds in both our 
TE and TP models) and 
Condorcet is vulnerable.  Our results are:

\begin{theorem}
\label{thm:partition-candidates}
Approval is 
vulnerable/certifiably-vulnerable
to constructive control by 
partition of candidates 
and run-off partition of candidates
in model TE and immune to 
constructive control by 
partition of candidates and run-off partition
of candidates in model TP\@.
Plurality, Condorcet, and 
approval are, in models TE and TP, 
respectively resistant, 
immune, and immune 
to destructive control
by partition of candidates and by run-off partition of candidates.
\end{theorem}

So Condorcet, though vulnerable to 
constructive control, is immune to destructive control here.
And, perhaps more interesting,
for constructive 
control, approval changes from vulnerable to immune
depending on the tie-handling rule.

We now turn to control of the voter set.  The intuition behind seeking
destructive control by adding or deleting voters is clear, e.g.,
getting out the vote and vote suppression.  We handle these two cases
together as their results are identical.

\subsubsection*{Control by Adding Voters}

\begin{description}
\item[Given:] A set of candidates $C$ and a distinguished candidate $c \in C$,
  a set $V$ of registered voters, an additional set $W$ of yet
  unregistered voters (both $V$ and $W$ have preferences over~$C$), and a
  positive integer $k \leq ||W||$.
\item[Question (constructive):] Is there a set of $k$ or fewer voters from $W$
  whose registration would assure that $c$ is the unique winner?
\item[Question (destructive):] Is there a set of $k$ or fewer voters from $W$
  whose registration would assure that $c$ is not the unique winner?
\end{description}

\subsubsection*{Control by Deleting Voters}

\begin{description}
\item[Given:] A set of candidates~$C$, a distinguished candidate $c \in C$, a
  set $V$ of voters, and a positive integer $k \leq ||V||$.
\item[Question (constructive):] Is there a set of $k$ or fewer voters in $V$
  whose disenfranchisement would assure that $c$ is the unique winner?
\item[Question (destructive):] Is there a set of $k$ or fewer voters in $V$
  whose disenfranchisement would assure that $c$ is not the unique
  winner?
\end{description}

Here, BTT92 shows that
for constructive control 
plurality is vulnerable
and 
Condorcet is resistant.  
Our results are:

\begin{theorem}
\label{thm:adding-deleting-voters}
Approval is 
resistant
to constructive control by 
adding voters and by deleting voters.
Plurality, Condorcet, and 
approval are all
vulnerable/certifiably-vulnerable
to destructive control
by adding voters and by deleting voters.
\end{theorem}

So Condorcet and approval, though resistant to 
constructive control, are vulnerable to destructive control here.

The final problem here results in a 
surprise.

\subsubsection*{Control by Partition of Voters} 

\begin{description}
\item[Given:] A set of candidates~$C$, a distinguished candidate $c \in C$,
  and a set $V$ of voters.
\item[Question (constructive):] Is there a partition of $V$ into $V_1$ and
  $V_2$ such that $c$ is the unique winner in the hierarchical two-stage
  election in which the survivors of $\parpair{C,V_1}$ and $\parpair{C,V_2}$
  run against each other with voter set~$V$?
\item[Question (destructive):] Is there a partition of $V$ into $V_1$
  and $V_2$ such that $c$ is not the unique winner in the hierarchical
  two-stage election in which the survivors of $\parpair{C,V_1}$ and
  $\parpair{C,V_2}$ run against each other with voter set~$V$?
\end{description}

In this last type of control, the voter set is partitioned into two
``subcommittees'' that both separately select their ``nominees,'' who
run against each other in the final decision stage.
Unlike BTT92,
we again distinguish between the two models Ties-Eliminate and
Ties-Promote defined above.  
That is, in the Ties-Eliminate model, if
two or more candidates tie for winning in a subcommittee's election,
no candidate is nominated by that subcommittee.  In contrast, in the
Ties-Promote model, all the candidates who tie for winning in a
subcommittee's election are nominated to run in the final decision
stage.  

We mention that both of our two tie-handling models, TE and TP,
differ from the model adopted in BTT92, where
they for vulnerability results about this problem adopt a third model in which
ties are handled not by a tie-handling rule but rather by changing the
decision problem itself to require the chair to find a partition that
completely avoids ties in any subcommittee.
We find our model the more natural, but for completeness we mention
that they obtained for this case, in their tie model,
a constructive-control vulnerability result 
for plurality.  For Condorcet and constructive control, BTT92
proves that resistance holds.
Our results are:

\begin{theorem}
\label{thm:partition-voters}
Approval is 
resistant
to constructive control by 
partition of voters 
in models TE and TP\@, and
vulnerable/certifiably-vulnerable
to destructive control by 
partition of voters 
in  models TE and TP\@.
Plurality is 
vulnerable/certifiably-vulnerable
to 
both constructive and destructive control
by partition of voters in model TE, and is 
resistant to 
both constructive and destructive control
by partition of voters in model TP\@.
Condorcet is 
vulnerable/certifiably-vulnerable
to 
destructive control
by partition of voters.
\end{theorem}

The most striking behavior here is that plurality 
voting varies between being vulnerable and being 
resistant, depending on the tie-handling rule.  
The loose  intuition for this is that in 
TE, at most one candidate wins each subcommittee and in polynomial
time we can explore every way this can happen.  In contrast,
under TP potentially any subset of candidates may move forward,
and in this particular setting, that flexibility is 
enough to support NP-completeness.
Also interesting is that both Condorcet
and approval, while resistant to constructive 
control, are vulnerable to destructive control.

\section{Proofs}
\label{sec:proofs}

In this section, we provide the proofs of the results stated in
Section~\ref{sec:results}.  Table~\ref{tab:overview-of-proofs} presents, for
each of the seven control types considered, the corresponding main result from
Section~\ref{sec:results} as well as the specific theorems, corollaries, and
examples from which this main result follows.

\begin{table*}[t]
\footnotesize
\centering
\begin{tabular}{||l|l|l||}
\hline\hline
                    & Main Result &              \\ 
Control by          & Stated as   & Follows from \\ \hline \hline
Adding Candidates   & Thm.~\ref{thm:adding-candidates} & 
 Thm.~\ref{thm:8706:hinged-absolute-corollaries},
 Cor.~\ref{cor:warp},
 Thm.~\ref{thm:vulnerable-destructive-addingcandidates-Condorcetapproval},
 Cor.~\ref{cor:resistance-destructive-addingcandidates-plurality}
                    \\ \hline
Deleting Candidates & Thm.~\ref{thm:deleting-candidates} &
 Thm.~\ref{thm:8706:hinged-absolute-corollaries},
 Cor.~\ref{cor:warp},
 Example~\ref{exa:8708:destructive-deletingpartitioncandidates-plurality-not-immune},
 \\ & & 
 Thm.~\ref{thm:vulnerable-constructive-deletingpartitioncandidates-approval},
 Cor.~\ref{cor:resistance-destructive-deletingcandidates-plurality}
 \\ \hline
Partition & Thm.~\ref{thm:partition-candidates} &
 Cor.~\ref{cor:warp},
 Thm.~\ref{thm:constructive-partition-approval-immunity},
  Example~\ref{exa:8707:constructive-deletingpartitioncandidates-approval-not-immune},
 \\ of Candidates & & 
Example~\ref{exa:8708:destructive-deletingpartitioncandidates-plurality-not-immune},
 Thm.~\ref{thm:vulnerable-constructive-deletingpartitioncandidates-approval},
 Cor.~\ref{cor:resistance-destructive-partitioncandidates-plurality}
 \\ \hline
Run-off Partition & Thm.~\ref{thm:partition-candidates} &
 Cor.~\ref{cor:warp},
 Thm.~\ref{thm:constructive-partition-approval-immunity},
  Example~\ref{exa:8707:constructive-deletingpartitioncandidates-approval-not-immune},
 \\ of Candidates & & 
 Example~\ref{exa:8708:destructive-deletingpartitioncandidates-plurality-not-immune},
 Thm.~\ref{thm:vulnerable-constructive-deletingpartitioncandidates-approval},
 Cor.~\ref{cor:resistance-destructive-run-off-partitioncandidates-plurality}
 \\ \hline
Adding Voters       & Thm.~\ref{thm:adding-deleting-voters} & 
Example~\ref{exa:8703:destructive-voterpartion-TE-TP-approval-not-immune},
 Example~\ref{exa:8704:destructive-voteradditiondeletion-pluralityCondorcet-not-immune},
  Example~\ref{exa:8705:destructive-voteradditiondeletion-approval-not-immune},
 \\ & & 
 Thm.~\ref{thm:vulnerable-destructive-voteradditiondeletion-plurality},
 Thm.~\ref{thm:vulnerable-destructive-voteradditiondeletion-Condorcet},
 Thm.~\ref{thm:vulnerable-destructive-voteradditiondeletion-approval},
 Thm.~\ref{thm:resistance-constructive-addingvoters-approval}
 \\ \hline
Deleting Voters     & Thm.~\ref{thm:adding-deleting-voters} &
Example~\ref{exa:8701:constructive-destructive-voterpartition-plurality-not-immune},
Example~\ref{exa:8702:destructive-voterpartion-Condorcet-not-immune},
  Example~\ref{exa:8703:destructive-voterpartion-TE-TP-approval-not-immune},
  Example~\ref{exa:8705:destructive-voteradditiondeletion-approval-not-immune},
 \\ & & 
 Thm.~\ref{thm:vulnerable-destructive-voteradditiondeletion-plurality},
 Thm.~\ref{thm:vulnerable-destructive-voteradditiondeletion-Condorcet},
 Thm.~\ref{thm:vulnerable-destructive-voteradditiondeletion-approval},
 Thm.~\ref{thm:resistance-constructive-deletingvoters-approval}

\\ \hline
Partition of Voters & Thm.~\ref{thm:partition-voters} & 
Example~\ref{exa:8701:constructive-destructive-voterpartition-plurality-not-immune},
Example~\ref{exa:8702:destructive-voterpartion-Condorcet-not-immune},
  Example~\ref{exa:8703:destructive-voterpartion-TE-TP-approval-not-immune},
Example~\ref{exa:8709:constructive-destructive-voterpartition-TP-plurality-not-immune},
 \\ & & 
  Example~\ref{exa:8710:constructive-addingdeletingvoterpartition-approval-not-immune},
 Thm.~\ref{thm:vulnerable-constructive-voterpartition-plurality},
 Thm.~\ref{thm:vulnerable-destructive-voterpartition-plurality},
 Thm.~\ref{thm:vulnerable-destructive-voterpartition-condorcet},
 \\ & & 
 Thm.~\ref{thm:vulnerable-destructive-voterpartition-approval},
 Cor.~\ref{cor:resistance-constructivedestructive--partitionvoters-TP-plurality},
 Thm.~\ref{thm:resistance-constructive-partitionvoters-TP-approval},
 Thm.~\ref{thm:resistance-constructive-partitionvoters-TE-approval}
\\ \hline\hline
\end{tabular}
\caption{Overview of results yielding the main results.  (For 
completeness, examples/theorems 
needed to establish susceptibility are listed even when they 
are invoked within a listed vulnerability or resistance theorem/corollary.)
\label{tab:overview-of-proofs}
}
\end{table*}
We first present the immunity and susceptibility
results.
Then we present the vulnerability
results, and finally we present 
the resistance results.  The proof techniques employed
range from 
political-science-axiom-fueled
arguments (for proving immunity), to
designing efficient algorithms (for proving vulnerability), to the
construction of $\np$-hardness reductions (for proving resistance).

\subsection{Proving Immunity and Susceptibility}\label{sec:immune-and-others}
For each of the 39 boldfaced entries in Table~\ref{tab:results},
this section must establish immunity if the entry is a boldface
``I'' and must establish susceptibility if the entry 
is a boldface ``R'' or a boldface ``V.''  (Recall that the definitions
of resistance and vulnerability require susceptibility, and so 
proving susceptibility
is a first step toward proving 
resistance or vulnerability.)

\subsubsection{Links Between Susceptibility Cases}\label{sss:links}
Rather than hand-proving each of the 39
cases just mentioned, it makes
sense to extract connections between the cases.  We start by stating
four easy but useful dualities.

\begin{theorem}\label{thm:two-way}
\begin{enumerate}
\item A voting system is 
susceptible to constructive control
by adding candidates if and only if it is 
susceptible to destructive control
by deleting candidates.
\item \label{part:two-way:two} A voting system is 
susceptible to constructive control
by deleting candidates if and only if it is 
susceptible to destructive control
by adding candidates.

\item A voting system is 
susceptible to constructive control
by adding voters if and only if it is 
susceptible to destructive control
by deleting voters.
\item A voting system is 
susceptible to constructive control
by deleting voters if and only if it is 
susceptible to destructive control
by adding voters.
\end{enumerate}
\end{theorem}
This theorem is easy to see, and so its proof is omitted.

We also have the following four implication results.
\begin{theorem}\label{thm:one-way}
\begin{enumerate}
\item \label{part:one-way:surprise} If a voting system is 
susceptible to constructive control
by partition of voters 
(in model TE or TP), then it is 
susceptible to constructive control
by deleting 
candidates.

\item \label{part:one-way:boring} If a voting system is 
susceptible to constructive control
by partition or run-off partition of candidates
(in model TE or TP), then it is 
susceptible to constructive control
by deleting candidates.

\item \label{part:one-way:three} If a voting system is 
susceptible to constructive control
by partition of voters 
in model TE, then it is 
susceptible to constructive control
by deleting 
voters.

\item \label{part:one-way:four} If a voting system is 
susceptible to destructive control
by partition or run-off partition of candidates
(in model TE or TP), then it is 
susceptible to destructive control
by deleting candidates.
\end{enumerate}
\end{theorem}

\begin{proofs}
We first prove parts~\ref{part:one-way:surprise} 
and~\ref{part:one-way:boring}.
Let $(C,V)$ be an election and $c \in C$ a candidate such that
$c$ is not the unique winner of $(C,V)$ and such that $c$
can be made the unique winner by partition of candidates, run-off
partition of candidates, or partition of voters.  Fix a partitioned
election such that $c$ is the unique winner of this election and
let $D \subseteq C$ be the set of candidates that participate
in the final round of the partitioned election.  Then $c$ is
the unique winner of $(D,V)$.  Thus, the voting system
is susceptible to constructive control by deleting candidates.

For part~\ref{part:one-way:three}, let
$(C,V)$ be an election and $c \in C$ a candidate such that
$c$ is not the unique winner of $(C,V)$ and such that $c$
can be made the unique winner by partition of voters in model TE\@.
Let $(V_1, V_2)$ be a partition of $V$ such that $c$ is the unique
winner of the partitioned election. 
Since $c$ makes it to the final run-off, and since we are
in model TE, 
$c$ is the unique winner of one of $(C,V_1)$ and~$(C,V_2)$.
Without loss of generality, suppose $c$ is the unique winner of~$(C,V_1)$.
Thus, the voting system
is susceptible to constructive control by deleting voters.

For part~\ref{part:one-way:four}, let
$(C,V)$ be an election and $c \in C$ a candidate such that
$c$ is the unique winner of $(C,V)$ and such that $c$
can be made to be not the unique winner by partition of candidates or run-off
partition of candidates.  Fix a partitioned
election such that $c$ is not the unique winner of this election and
let $D \subseteq C$ be the set of candidates that participate
in the final round of the partitioned election. 
If $c \in D$, then $c$ is not the unique winner of $(D,V)$.
If $c \not \in D$, then  $c$ is not the unique winner of
the subelection involving $c$.  Thus, the voting system
is susceptible to destructive control by deleting candidates.~\end{proofs}

Let us say that a voting system is \emph{voiced} if in any
election that has exactly one candidate, that candidate is always a
(and thus, the unique) winner.  
Note that 
plurality, Condorcet, and approval are all voiced systems.
For voiced systems, we have the following 
three additional results.

\begin{theorem}\label{thm:hinged}
\begin{enumerate}
\item \label{part:hinged:one} If a voiced voting system is 
susceptible to destructive control
by partition of voters 
(in model TE or TP), then it is 
susceptible to destructive control
by deleting 
voters.

\item \label{part:hinged:two} Each voiced voting system is 
susceptible to constructive control
by deleting 
candidates.

\item \label{part:hinged:three} Each voiced voting system is 
susceptible to destructive control
by adding
candidates.
\end{enumerate}
\end{theorem}

\begin{proofs}
Fix a voiced voting system.

For part~\ref{part:hinged:one}, suppose that our voting system
is immune to destructive control by deleting voters.
We will show that it is also
immune to destructive control by partition of voters.
Let $(C,V)$ be an election such that $c$ is the unique winner
of $(C,V)$, and let $(V_1, V_2)$ be an arbitrary partition of
$V$.   Then $c$ is the unique winner of $(C,V_1)$
and of $(C,V_2)$,
and so $c$ is the only candidate participating in the final run-off
(both in model TP and in model TE).  Since the voting system
is voiced, $c$ wins the final run-off, and is thus the unique winner
of the partitioned election.  It follows that the voting system is
immune to destructive control by partition of voters.

For part~\ref{part:hinged:two}, let $C = \{c,d\}$ and let $V$ be
an arbitrary set of voters with preferences over $C$.
At least one of the candidates
is not a unique winner of $(C,V)$.  Without loss of generality,
let $c$ not be a unique winner of $(C,V)$.  Since the voting system
is voiced, $c$ is the unique winner of $(\{c\},V)$. It follows that
the voting system is susceptible to constructive control
by deleting candidates.

Part~\ref{part:hinged:three} follows immediately
from part~\ref{part:hinged:two} of this theorem and
part~\ref{part:two-way:two} of Theorem~\ref{thm:two-way}.~\end{proofs}

Since
plurality, Condorcet, and approval are all voiced systems, we 
immediately have from 
Theorems~\ref{thm:hinged} and~\ref{thm:two-way} the following
results that yield susceptibility results for four of
Table~\ref{tab:results}'s boldface ``R'' and boldface ``V'' 
entries.

\begin{theorem}\label{thm:8706:hinged-absolute-corollaries}
Plurality, Condorcet, and approval are each susceptible to 
destructive control by adding candidates.  
Approval is susceptible to 
constructive control by deleting candidates.  
\end{theorem}

\subsubsection{Immunity Results}

We start by proving the immunity results of
Theorems~\ref{thm:adding-candidates}, \ref{thm:deleting-candidates},
and~\ref{thm:partition-candidates}.  These results are generally clear
from the definitions.  
Bartholdi, Tovey, 
and Trick~\cite{bar-tov-tri:j:control}
observed\footnote{Their paper is somewhat nonspecific regarding the uniqueness
issue and merely says WARP.
}
that immunity to constructive control by adding candidates
follows from the
``unique'' version of the Weak Axiom of Revealed Preference (denoted by
Unique-WARP), which says that a unique winner among a collection of
candidates always remains a unique winner among every subcollection of
candidates that includes him/her.  
Theorem~\ref{thm:warp-new}
states ways in which Unique-WARP influences 
a  variety of destructive control scenarios.

\begin{theorem}
{\bf\cite{bar-tov-tri:j:control}}\label{thm:warp-old}~$\,$Any 
voting system that satisfies 
Unique-WARP
is immune to constructive control by
adding candidates.
\end{theorem}
\begin{theorem}\label{thm:warp-new}
Any voting system that satisfies 
Unique-WARP
is immune to destructive control by deleting
candidates and (in both model TE and model TP)
to destructive control by partition and run-off partition of candidates.
\end{theorem}

Theorem~\ref{thm:warp-new} follows from
Theorem~\ref{thm:warp-old} via 
Theorems~\ref{thm:two-way} and~\ref{thm:one-way} (and also 
is directly clear).

Bartholdi, Tovey, and Trick~\cite{bar-tov-tri:j:control} note that
Theorem~\ref{thm:warp-old} can be applied to show that Condorcet voting is
immune to constructive control by adding candidates.  We state further
immunity results, via Theorems~\ref{thm:warp-old} and~\ref{thm:warp-new}, 
as Corollary~\ref{cor:warp}.

\begin{corollary}
\label{cor:warp}
\begin{enumerate}
\item Condorcet voting is immune to destructive control by deleting
candidates, partition of candidates, and run-off partition of
candidates.

\item Approval voting is immune to constructive control by adding
candidates, and is immune to destructive control by deleting
candidates and by partition and run-off partition of candidates (in both
the TE and the TP models).
\end{enumerate}
\end{corollary}

\begin{proofs}
Both Condorcet and approval voting clearly satisfy Unique-WARP.
The result now follows from
Theorems~\ref{thm:warp-old} and~\ref{thm:warp-new}.~\end{proofs}

Note that, unlike Condorcet and approval, plurality voting does not
satisfy Unique-WARP, and we will see that
immunity does not hold for plurality in
any control scenario considered here.

We now state and prove the final two immunity results.

\begin{theorem}
\label{thm:constructive-partition-approval-immunity}
Approval voting is immune to constructive control
by partition and run-off partition of candidates in model~TP\@.
\end{theorem}

\begin{proofs}
It is easy to see that in approval voting, a candidate $c$
is the unique winner if and only if there is a unique
candidate with a maximum number of Yes votes and $c$ is 
that candidate.  In the TP model, this remains true  even under 
the two partitioning schemes.~\end{proofs}

\subsubsection{Susceptibility Results}\label{sss:susceptibility}
We now turn to proving susceptibility results for the 28 boldface
``R'' and boldface ``V'' boxes in Table~\ref{tab:results}.  

Note that in concept each of the ``R'' and ``V'' claims made by
Bartholdi, Tovey, and Trick~\cite{bar-tov-tri:j:control} is 
asserting a susceptibility result, and around eight of those---via our
Section~\ref{sss:links} theorems---imply eight of the 28
susceptibility
results that we need.  However, Bartholdi, Tovey, and
Trick~\cite{bar-tov-tri:j:control} generally do not prove
their susceptibility claims, and so 
we will prove susceptibility here for all our 28 cases.

Now, how does one prove susceptibility?  One need simply give an
example in each case.  Alternatively, some cases we get indirectly
from an earlier example via our Section~\ref{sss:links} theorems.
However, please note that even in those cases, there is implicitly a
concrete example, as the theorems of Section~\ref{sss:links} have
constructive proofs.  Also, we mention again that plurality,
Condorcet, and approval are all voiced voting systems (in the sense
defined in Section~\ref{sss:links}).

We first show that plurality is not immune to constructive or
destructive control by partition of voters in model~TE or 
to destructive control by 
deleting voters.

\begin{example}\label{exa:8701:constructive-destructive-voterpartition-plurality-not-immune}
Let us consider
destructive control by partition of voters in model~TE\@.
Let $C = \{a,b,c\}$, and define $V$ to consist of five voters with
the following preferences:
\[
\begin{array}{l@{\hspace*{4mm}}l}
v_1 : a > b > c, & v_2 : a > b > c, \\
v_3 : b > a > c, & v_4 : b > a > c, \\
v_5 : c > a > b. &
\end{array}
\]
Thus, $c$ does not win in~$\parpair{C,V}$.  However, if we partition
$V$ into $V_1 = \{v_5\}$ and $V_2 = V - V_1$, $c$ trivially wins the
subelection~$\parpair{C,V_1}$, but $a$ and $b$ tie for winner in the
subelection~$\parpair{C,V_2}$, so none of them proceeds to the
run-off with $c$ in model~TE\@.  It follows that $c$ is the unique
run-off winner.  Thus, plurality voting is susceptible to constructive
control by partition of voters in model~TE\@.

For the destructive case, $a$ is the unique plurality winner in the
election $\parpair{C,V'}$, where $V' = \{v_1, v_2, v_3, v_5\}$.  Now,
partitioning $V'$ into $V_{1}^{'} = \{v_1, v_3\}$ and $V_{2}^{'} =
\{v_2, v_5\}$ implies that none of the two subcommittees nominates a
candidate in model~TE, due to ties.  In particular, $a$ is not the
unique run-off winner, and plurality voting thus is
susceptible to destructive control by partition of voters in
model~TE\@.

By Theorem~\ref{thm:hinged}, this latter susceptibility claim implies
that plurality is susceptible to destructive control by deleting
voters (which is another of the 28 boldfaced ``R''-or-``V'' boxes 
we are handling). 

\end{example}

We now prove that Condorcet voting is
susceptible to destructive control by partition
of voters and to destructive control by deleting voters.

\begin{example}
\label{exa:8702:destructive-voterpartion-Condorcet-not-immune}
Let us consider 
destructive control by partition
of voters.
Let $C = \{a,b,c\}$, and define $V$ to consist of seven voters with
the following preferences:
\[
\begin{array}{l@{\hspace*{4mm}}l@{\hspace*{4mm}}l}
v_1 : c > a > b, & v_2 : c > a > b, & v_3 : c > b > a, \\
v_4 : b > a > c, & v_5 : b > c > a, & \\
v_6 : a > b > c, & v_7 : a > c > b. &
\end{array}
\]
Since in pairwise contests four voters prefer $c$ to $a$ and four
voters prefer $c$ to~$b$, $c$ is the Condorcet winner in the
election~$\parpair{C,V}$.  However, partitioning $V$ into $V_1 =
\{v_1, v_7\}$ and $V_2 = V - V_1$ implies that there is no Condorcet
winner in the subelection $\parpair{C,V_1}$, and $b$ is the
Condorcet winner in the subelection $\parpair{C,V_2}$.  
Thus, Condorcet voting is susceptible to destructive control
 by partition of voters.

By Theorem~\ref{thm:hinged}, this susceptibility claim implies
that Condorcet is also susceptible to 
destructive control by deleting voters.

\end{example}

We now prove that approval voting is susceptible
to destructive control by partition of voters in models TE and TP,
to destructive control by deleting voters,
and to constructive control by adding voters.

\begin{example}
\label{exa:8703:destructive-voterpartion-TE-TP-approval-not-immune}
Let $C = \{a,b,c\}$, and define $V$ to consist of the following ten
voters (specified by vectors from $\{0,1\}^3$, with the first,
second, and third bits 
specifying approval/disapproval for $a$, $b$, and $c$): 
$v_1 = v_2 = v_3 = v_4 = 001$, 
$v_5 = v_6 = v_7 = 100$, and 
$v_8 = v_9 = v_{10} = 010$.  
In $\parpair{C,V}$, $c$ is the unique approval winner.  But if $V$ is
partitioned into $V_1 = \{v_1, v_2, v_5, v_6, v_7\}$ and $V_2 = V -
V_1$, then $a$ and $b$ are nominated by the subcommittees $V_1$
and~$V_2$, respectively, and tie for winner in the
run-off.  Thus, approval voting is susceptible to 
destructive control by partition of voters, both in model TE and 
in TP\@.

By Theorem~\ref{thm:hinged}, this susceptibility claim implies
that approval voting is also susceptible to 
destructive control by deleting voters.
And that claim itself,
by Theorem~\ref{thm:two-way}, implies
that approval voting is also susceptible to 
constructive control by adding voters.
\end{example}

The
following example shows that both plurality voting and Condorcet
voting are not immune to destructive control by adding
voters.

\begin{example}
\label{exa:8704:destructive-voteradditiondeletion-pluralityCondorcet-not-immune}
Let $C = \{a,b,c\}$.  Define $V$ to consist of one registered voter
$v$ with preference $c > a > b$, and define $W$ to consist of one as yet
unregistered voter $w$ with preference $a > c > b$.  Candidate $c$ is the
unique winner---both for plurality and Condorcet voting---in the
election~$\parpair{C,V}$, yet registration of $w$ would assure that
$a$ and $c$ tie in first-place votes in~$\parpair{C,V \cup W}$, so $c$
is not the unique plurality winner of this election.  Similarly, $c$
is no longer the Condorcet winner in~$\parpair{C,V \cup W}$.
Thus, both plurality and Condorcet voting are susceptible to
destructive control 
by adding voters.
\end{example}

The following example shows that approval voting is not immune to
destructive control by adding 
voters or to constructive control by deleting voters.

\begin{example}
\label{exa:8705:destructive-voteradditiondeletion-approval-not-immune}
Let us consider
destructive control by adding voters.
Let $C = \{a,b,c\}$.  Define $V$ to consist of one registered voter $v
= 001$ (i.e., $v$ approves of $c$ and disapproves of $a$ and $b$), 
and define $W$ to consist
of one unregistered voter $w = 100$.  In $\parpair{C,V}$, $c$ is the unique
approval winner, yet registration of $w$ would assure that
$a$ and $c$ tie for winner in~$\parpair{C,V \cup W}$, so $c$
is not the unique plurality winner of this election.
Thus, approval voting is susceptible to destructive control 
by
adding voters.

By Theorem~\ref{thm:two-way}, this susceptibility claim implies
that approval voting is also susceptible to 
constructive control by deleting voters (and indeed, as implicit
in the proof of 
Theorem~\ref{thm:two-way}, this very same example works to show that).
\end{example}

We now show susceptibility for approval voting
to 
constructive control by
partition of candidates and run-off partition of
candidates, both in model~TE\@.  

\begin{example}
\label{exa:8707:constructive-deletingpartitioncandidates-approval-not-immune}
Let $C = \{a, b,c\}$ be the candidate set.
Let the voter set $W$ consist of
the two voters with vector representation $w_1 = 111$ and $w_2 = 110$,
respectively.  Then $c$ loses to both $a$ and~$b$, who tie for winning in
the election~$\parpair{C,W}$.  But if we partition $C$ into $C_1 =
\{a,b\}$ and $C_2 = \{c\}$, then no one moves forward from the
subelection $\parpair{C_1,W}$ in model~TE, so $c$ wins overall.
The same example works for the run-off partition of candidates case,
since no one moves forward from the
subelection $\parpair{C_1,W}$ in model~TE and $c$ first wins the
subelection $\parpair{C_2,W}$ and then the run-off.
Thus, approval voting is susceptible to constructive control by
both partition of candidates in model~TE and run-off
partition of candidates in model~TE\@.
\end{example}

Example~\ref{exa:8708:destructive-deletingpartitioncandidates-plurality-not-immune}
shows that plurality voting is susceptible to destructive control by
partition and run-off partition of candidates (both
in model TE and~TP), and to destructive control by deleting 
candidates.

\begin{example}
\label{exa:8708:destructive-deletingpartitioncandidates-plurality-not-immune}
Let us consider the partition cases.
Let $C = \{a, b, c, d\}$ be the candidate set, and define the voter set $V$ to
consist of the following seven voters:
\begin{itemize}
\item $3$ voters of the form $c > a > b > d$,
\item $2$ voters of the form $a > d > b > c$, and
\item $2$ voters of the form $b > d > a > c$.
\end{itemize}
Note that $c$ is the unique plurality winner in the election~$\parpair{C,V}$.

Now, partition the candidate set $C$ into $C_1 = \{a,c\}$ and~$C_2 = \{b,d\}$.
Then $a$ is the unique plurality winner in the subelection~$\parpair{C_1,V}$.
So $c$ is dethroned in the partition of candidates setting.
$C_1 = \{a,c\}$ and~$C_2 = \{b,d\}$ also dethrones $c$ in the
run-off partition of candidates setting.
Thus, plurality voting is susceptible to destructive control by partition and by
run-off partition of candidates.  Since each subelection has a unique winner
($a$ in~$\parpair{C_1,V}$ and $b$ in~$\parpair{C_2,V}$), this is true
regardless of the tie-handling rule.

By Theorem~\ref{thm:one-way}, these susceptibility claims imply
that plurality voting is also susceptible to 
destructive control by deleting candidates.

\end{example}

The next example shows that plurality voting is susceptible to constructive
and destructive control by partition of voters in model~TP\@.

\begin{example}
\label{exa:8709:constructive-destructive-voterpartition-TP-plurality-not-immune}
Let $C = \{a, b, c\}$ be the candidate set, and define the voter set $V$ to
consist of the following eight voters:
\begin{itemize}
\item $3$ voters (say $u_1$, $u_2$, and~$u_3$) of the form $a > c > b$,
\item $2$ voters (say $v_1$ and $v_2$) of the form $b > a > c$, and
\item $3$ voters (say $w_1$, $w_2$, and~$w_3$) of the form $c > a > b$.
\end{itemize}

For the constructive case, note that $c$ is not the unique plurality winner in
the election~$\parpair{C,V}$, since $a$ and $c$ are tied for first place.
Now, partition $V$ into $V_1 = \{u_1, u_2, w_1, w_2, w_3\}$ and $V_2 = \{u_3,
v_1, v_2\}$.  Then $c$ is the unique plurality winner in the
subelection~$\parpair{C,V_1}$, $b$ is the unique plurality winner in the
subelection~$\parpair{C,V_2}$, and $c$ wins the run-off against~$b$.  Thus,
plurality voting is not immune to constructive control by partition of voters
in model~TP\@.

For the destructive case, consider the election $\parpair{C,V'}$ with $V' = V
\cup \{v_3,w_4\}$, where $v_3$ votes $b > a > c$ and $w_4$ votes $c > a > b$.
In $\parpair{C,V'}$, $c$ is the unique plurality winner.  Partition $V'$ into
$V'_{1} = \{u_1, u_2, u_3, w_1, w_2\}$ and $V'_{2} = \{v_1, v_2, v_3, w_3,
w_4\}$.  Then $a$ is the unique plurality winner of the
subelection~$\parpair{C,V'_{1}}$, $b$ is the unique plurality winner of the
subelection~$\parpair{C,V'_{2}}$, and $a$ wins the run-off against~$b$.  So
$c$ is dethroned.  Thus, plurality voting is not immune to destructive control
by partition of voters in model~TP\@.
\end{example}

Finally, we show that approval voting is susceptible to constructive control
by 
partition of voters in models TE
and~TP\@.

\begin{example}
\label{exa:8710:constructive-addingdeletingvoterpartition-approval-not-immune}
Let $C = \{a,b,c\}$ be the candidate set.  
Define the voter set $V$ to
consist of the following eight voters: $v_1 = v_2 = v_3 = 100$, $v_4 = v_5 =
010$, and $v_6 = v_7 = v_8 = 001$.  In $\parpair{C,V}$, $a$ and $c$ are tied.
Now, partition $V$ into $V_1 = \{v_1, v_2, v_6, v_7, v_8\}$ and $V_2 = \{v_3,
v_4, v_5\}$.  Candidate $c$ is the unique approval winner in the
subelection~$\parpair{C,V_1}$, $b$ is the unique approval winner in the
subelection~$\parpair{C,V_2}$, and $c$ wins the run-off against~$b$.  This
works both in model TE and~TP, since ties do not occur in the subelections in our 
construction.
So approval voting is susceptible to constructive control by
partition of voters (both in TE and~TP).
\end{example}

\subsection{Proving Vulnerability}
\label{sec:vulnerability-proofs}

The certifiably-vulnerable results (which here imply the vulnerable
results) range from clear greedy algorithms to trickier algorithms
based on characterizing the ways in which a candidate can be made to win
(in the constructive case) or can be precluded
from winning (in the destructive case).  The
more surprising of these have to do with the tie-handling cases of
partition problems---where the chair can at times do shrewd things
(e.g., shift voters counterintuitively to induce ties that kill off
stronger candidates).

\subsubsection{Partition of Voters}

We start with the ``control by partition of voters'' problems.  For
plurality voting, we here obtain the same results in the constructive
and the destructive case, as stated in Table~\ref{tab:results} and in
Theorem~\ref{thm:partition-voters}.  On the other hand, the question
of whether resistance or vulnerability holds depends on which
tie-handling rule is chosen.

\begin{theorem}
\label{thm:vulnerable-constructive-voterpartition-plurality}
In model~TE, plurality voting is vulnerable/certifiably-vulnerable to
constructive control by partition of voters.
\end{theorem}

\begin{proofs}
By Example~\ref{exa:8701:constructive-destructive-voterpartition-plurality-not-immune},
susceptibility holds.

Given a set of candidates~$C$, a distinguished candidate $c \in C$,
and a voter set~$V$, we describe a polynomial-time algorithm for this
problem.  For any partition $\parpair{V_1,V_2}$ of the voter set~$V$,
let $\nominee{C,V_i}$, $i \in \{1,2\}$, denote the set of candidates
who are nominated by the subcommittee $V_i$ (with candidates~$C$) for
the run-off in model~TE\@.  To ensure that $c$ is the unique winner,
under the desired partition setup, we may without loss of generality
focus on the following five cases (Cases~3 and~5 are not 
necessarily disjoint):
\begin{description}
\item[Case~1:] $\nominee{C,V_1} = \{c\}$ and $\nominee{C,V_2} =
\emptyset$ due to $V_2 = \emptyset$.

\item[Case~2:] $\nominee{C,V_1} = \{c\}$ and $\nominee{C,V_2} = \{c\}$.

\item[Case~3:] $\nominee{C,V_1} = \{c\}$ and $\nominee{C,V_2} =
\emptyset$ due to $c$ and $d$ (and possibly additional other
candidates) tying, where $c \neq d$.

\item[Case~4:] $\nominee{C,V_1} = \{c\}$ and $\nominee{C,V_2} = \{d\}$,
where $c \neq d$.

\item[Case~5:] $\nominee{C,V_1} = \{c\}$ and $\nominee{C,V_2} =
\emptyset$ due to $d$ and $e$ (and possibly additional other
candidates) tying, where $c \neq d \neq e \neq c$.
\end{description}

In Case~1, it clearly suffices to check whether $c$ is an overall
plurality winner.  Note further that if Case~2 holds for some
partition~$\parpair{V_1,V_2}$, then $c$ must be an overall plurality
winner, and thus will also win via the
partition~$\parpair{V,\emptyset}$.  We now argue that the same is true
in Case~3.  For any candidate~$i$, let $\score{i}$ denote the number of
voters who rank $i$ first-place in~$\parpair{C,V}$.
In Case~3, note that for all $e \in C - \{c\}$,
\begin{eqnarray*}
\score{e} < \score{c},
\end{eqnarray*}
since $c$ has strictly more first-place votes than $e$ in
$\parpair{C,V_1}$ and $e$ at best ties $c$ for first-place votes in
$\parpair{C,V_2}$. It follows that
$c$ must already be an overall plurality winner in Case~3, and thus
will also win via the partition~$\parpair{V,\emptyset}$.

So, our algorithm, after checking whether $c$ is a plurality winner
overall (thus catching Cases~1, 2, and~3), will by brute force check
whether Case~4 or Case~5 can be made to hold for some partition of the
voter set.

Given $C$, $c$, and $V$ as above, our polynomial-time algorithm
proceeds as follows.  If $c$ is a plurality winner of $\parpair{C,V}$,
output $\parpair{V,\emptyset}$ as a successful partition and halt;
else if $||C|| = 2$, then output ``control impossible'' (which
in this context means that making $c$ 
a unique winner is impossible) and
halt.
Otherwise, we first try to make
Case~4 hold and then, if that fails, try to make Case~5 hold.  These
two tests are implemented by the two loops described below, and if
they both fail, control is not possible.
\begin{description}
\item {\bf Loop trying to make Case~4 hold:} For each $d \in C$, $d
  \neq c$, such that $c$ beats $d$ in a pairwise plurality election by
  the voters in~$V$, do the following: If it holds that, for each $e
  \in C$ with $c \neq e \neq d$,
\begin{eqnarray*}
\score{e} \leq \score{c} + \score{d} - 2,
\end{eqnarray*}
then output $\parpair{V_1,V_2}$ as a successful partition and halt,
where $V_1$ consists of all $\score{c}$ voters whose first choice is $c$
and 
 exactly $\min(\score{e}, \score{c} - 1)$ of the voters whose first choice
 is~$e$, and where $V_2 = V - V_1$.

\item {\bf Loop trying to make Case~5 hold:} If the loop trying to
  make Case~4 hold was not successful, then for each $d \in C$ and for
  each $e \in C$ such that $||\{c,d,e\}|| = 3$
and $\score{d} \leq \score{e}$, do the
following:
If it holds that, for each $f \in C - \{c\}$,
\begin{eqnarray*}
\score{f} \leq \score{c} + \score{d} - 1,
\end{eqnarray*}
then output $\parpair{V_1,V_2}$ as a successful partition and halt,
where $V_1$ consists of all $\score{c}$ voters whose first choice is $c$,
of exactly $\score{e} - \score{d}$ of the voters whose first choice
is $e$, and for all $f \in C - \{c,d,e\}$,
exactly $\min(\score{f}, \score{c} - 1)$ of the voters
whose first choice is~$f$, and where $V_2 = V - V_1$.
\end{description}
Otherwise (i.e., if the Case~5 loop was not successful either), $c$
cannot win, so we output ``control impossible'' and halt.~\end{proofs}

We now make a general remark.
In various cases, our polynomial-time algorithms have loops.  In some
cases, these loops can be collapsed or removed.  Doing so
\begin{itemize}
\item improves the runtime and makes the algorithm look simpler, but
\item makes it a bit harder to see that the algorithm is correct.
\end{itemize}
Since correctness is what we most care about, we do not collapse such
loops.  But let us explicitly mention the ``look'' of such collapses.
In the proof of
Theorem~\ref{thm:vulnerable-constructive-voterpartition-plurality}
above, the ``For each $d \in C$, $d \neq c$, such that $c$ beats $d$
in a pairwise plurality election by the voters in~$V$, do\ldots''~loop
trying to make Case~4 hold in the algorithm can safely be changed to:
``If there exists some
$d' \in C$, $d' \neq c$, such that $c$ beats $d'$ in a pairwise
plurality election by the voters in~$V$, then let $d$ be some such
$d'$ for which $\score{d}$ is maximized among all such $\score{d'}$ and
do\ldots'' This is a legal loop collapse, since if some $d'$ works, then
it works for all $d''$ that can pairwise beat $c$ in a
run-off whose $\score{d''}$ is maximum.
Again, this is just an example,
and to have our correctness as unobscured as possible and as our focus
is on the gap between $\p$ and $\np$-hard, we in general forgo such
optimizations of the precise polynomial of the runtime.

We now turn to the destructive analog of
Theorem~\ref{thm:vulnerable-constructive-voterpartition-plurality}.

\begin{theorem}
\label{thm:vulnerable-destructive-voterpartition-plurality}
In model~TE, plurality voting is vulnerable/certifiably-vulnerable to
destructive control by partition of voters.
\end{theorem}

\begin{proofs} 
  That susceptibility holds in this case has been shown in
  Example~\ref{exa:8701:constructive-destructive-voterpartition-plurality-not-immune}.

Given a set of candidates~$C$, a distinguished candidate $c \in C$,
and a voter set~$V$, our polynomial-time algorithm for this control
problem works as follows.  If $C = \{c\}$, output ``control
impossible'' and halt, as $c$ must win; else if $c$ already is not the
unique plurality winner, output $\parpair{V,\emptyset}$ as a
successful partition and halt.
Now, we check if
every voter's first choice is $c$ or if $||C|| = 2$, and if one of
these two conditions is true, we output ``control impossible'' and
halt, since $c$ cannot help but win.

Again, let $\score{i}$ denote the number of voters who rank candidate $i$
first-place.  Let $d$ be a candidate who other than $c$ got the most
first-place votes, and let $e$ be a candidate who other than $c$ and $d$
got the most first-place votes.
We can certainly
dethrone $c$ if
\begin{eqnarray}
\label{equ:vulnerable-destructive-voterpartition-plurality-1}
\score{c} \leq \score{d} + \score{e} .
\end{eqnarray}
Namely, if
Equation~(\ref{equ:vulnerable-destructive-voterpartition-plurality-1})
holds, we output $\parpair{V_1,V_2}$ as a successful partition and
halt, where $V_1$ consists of all $\score{d}$ voters whose first choice is
$d$ and exactly $\score{d}$ voters whose first choice is $c$ (recall that in
the current case we already know that $\score{c} > \score{d}$), 
and where $V_2 = V - V_1$.  
Then $c$ and $d$ will tie for winner in $\parpair{C,V_1}$,
so no one will be nominated by the subcommittee $V_1$ in model~TE, 
and $e$ will tie or beat $c$ in $\parpair{C,V_2}$, so $c$ is not 
nominated by the subcommittee $V_2$ either.

On the other hand, if
Equation~(\ref{equ:vulnerable-destructive-voterpartition-plurality-1})
is not satisfied, we have
\[
\score{c} > \score{d} + \score{e},
\]
so in any
partition $\parpair{V_1,V_2}$, $c$ clearly will triumph in one of
$\parpair{C,V_1}$ or~$\parpair{C,V_2}$.
Thus, we now know it is
impossible to make sure that $c$ loses in both subcommittees.  If $c$
is nominated by both subcommittees (in model~TE), $c$ trivially is the
unique winner of the final run-off.  So, our algorithm now checks if
it is possible for $c$ to win in exactly one subcommittee, and yet can
be made to not be the unique winner of the final run-off.  For this to
happen, it is (given the case we are in) a necessary and sufficient
condition that there exists some candidate $d$ such 
that:
\begin{itemize}
\item $d \neq c$,
\item $d$ ties or beats $c$ in a pairwise plurality election, and
\item for each candidate~$e$, $c \neq e \neq d$, we have that $\score{e} <
\score{c} + \score{d} - 2$.
\end{itemize}
We can in polynomial time brute-force check whether the above three
conditions hold for some candidate~$d$, 
and if they do, let $d'$ be some such candidate $d$
and output $\parpair{V_1,V_2}$ as a successful partition and halt,
where $V_1$ consists of all $\score{c}$ voters whose first choice is $c$ and,
for each candidate $e$ with $c \neq e \neq d'$, of exactly 
$\min(\score{c} - 1, \score{e})$ 
voters whose first choice is~$e$, and where $V_2 = V - V_1$.
Finally, if the above two conditions cannot be satisfied for any~$d$,
output ``control impossible'' and halt.~\end{proofs}

We now prove that Condorcet voting is
vulnerable/certifiably-vulnerable to destructive control by partition
of voters.  

\begin{theorem}
\label{thm:vulnerable-destructive-voterpartition-condorcet}
Condorcet voting is vulnerable/certifiably-vulnerable to destructive
control by partition of voters.
\end{theorem}

\begin{proofs}
By Example~\ref{exa:8702:destructive-voterpartion-Condorcet-not-immune},
susceptibility holds.

Given a set of candidates~$C$, a distinguished candidate $c \in C$,
and a voter set~$V$, our polynomial-time algorithm for this control
problem proceeds in three stages:
\begin{enumerate}
\item {\bf Checking the trivial cases:} If $C = \{c\}$, output
``control impossible'' and halt, as $c$ must win.
Otherwise, if $c$ already
is not the Condorcet winner, output $\parpair{V,\emptyset}$ as a
successful partition and halt.  Otherwise, if $||C|| = 2$, output
``control impossible'' and halt, since in this case $c$ {\em
is\/} the Condorcet winner, so $c$ is preferred by a strict majority
of votes to the other candidate and thus will win at least one
subcommittee and also the run-off.

\item {\bf Loop:} Now, if none of the trivial cases applies, for each
$a, b \in C$ with $||\{a,b,c\}||=3$, we test whether we can make $a$
tie or beat $c$ in $\parpair{C,V_1}$ and make $b$ tie or beat $c$
in~$\parpair{C,V_2}$.  For each voter, we will now focus just on the
ordering of~$a$, $b$, and~$c$.  We use the following notation.  Denote
the number of voters with order $c > a > b$ or $c > b > a$ by~$W_c$,
with order $a > b > c$ or $b > a > c$ by~$L_c$, with order $a > c > b$
by~$S_a$, and with order $b > c > a$ by~$S_b$.
If $W_c - L_c > S_a + S_b$, then this $a$ and $b$ are hopeless, so
move on to consider the next $a$ and $b$ in the loop.  Otherwise, we
have
\begin{eqnarray}
\label{equ:vulnerable-destructive-voterpartition-condorcet-1}
W_c - L_c & \leq & S_a + S_b.
\end{eqnarray}
Output $\parpair{V_1,V_2}$ as a successful partition and halt, where
$V_1$ contains all the $S_a$ voters with order $a > c > b$, and also
$\min(W_c, S_a)$ voters contributing to~$W_c$, and where $V_2 = V - V_1$.

In $\parpair{C,V_1}$, $a$ ties or beats~$c$, since $a$ gets $S_a$
votes and $c$ gets $\min(W_c, S_a)$ votes.  And in $\parpair{C,V_2}$,
$b$ ties or beats~$c$, since there are $S_b + L_c$ voters who prefer
$b$ to~$c$, and there are $W_c - \min(W_c, S_a)$ voters who prefer $c$
to~$b$.  Thus, to prove that the construction works, we need that
\[
S_b + L_c \geq W_c - \min(W_c, S_a),
\]
which is equivalent to
\begin{eqnarray}
\label{equ:vulnerable-destructive-voterpartition-condorcet-2}
S_b + \min(W_c, S_a) & \geq & W_c - L_c.
\end{eqnarray} 
But if $S_a \leq W_c$ then
Equation~(\ref{equ:vulnerable-destructive-voterpartition-condorcet-2})
is implied by
Equation~(\ref{equ:vulnerable-destructive-voterpartition-condorcet-1}),
and if $S_a > W_c$ then
Equation~(\ref{equ:vulnerable-destructive-voterpartition-condorcet-2})
follows immediately from the fact that $S_b + L_c \geq 0$.  Thus, $b$
indeed ties or beats $c$ in $\parpair{C,V_2}$.

\item {\bf Termination:} If in no loop iteration did we find an $a$
and $b$ that allowed us to output a partition of voters
dethroning~$c$, then output ``control impossible'' and halt.
\end{enumerate}
This completes the proof of
Theorem~\ref{thm:vulnerable-destructive-voterpartition-condorcet}.~\end{proofs}

We now prove that approval voting is vulnerable/certifiably-vulnerable
to destructive control by partition of voters in models TE and TP\@.  

\begin{theorem}
\label{thm:vulnerable-destructive-voterpartition-approval}
Approval voting is vulnerable/certifiably-vulnerable to destructive
control by partition of voters in models TE and TP\@.
\end{theorem}

\begin{proofs}
  That susceptibility holds in this case is shown by
  Example~\ref{exa:8703:destructive-voterpartion-TE-TP-approval-not-immune}.

We describe two polynomial-time algorithms for these two control
problems, one for TE and one for TP\@.  Given a set of candidates~$C$,
a distinguished candidate $c \in C$, and a voter set~$V$, both
algorithms again
proceed in the following three phases:
\begin{enumerate}
\item {\bf Checking the trivial cases:} If $C = \{c\}$, output
``control impossible'' and halt, as $c$ must win.  Otherwise, if $c$ already
is not the unique winner, output $\parpair{V,\emptyset}$ as a
successful partition and halt.  Otherwise, if $||C|| = 2$, output
``control impossible'' and halt, since in this case
$c$ {\em
is\/} the unique winner, so $c$ will win in at least one subcommittee and
will also win the run-off.

\item {\bf Loop:} In this phase, if none of the trivial cases applies,
we try to find a pair of candidates, $a$ and~$b$, that allows us to
determine a successful partition of voters.  This phase is described
below, separately for TE and TP\@.

\item {\bf Termination:} If in no loop iteration did we find an $a$
and $b$ that allowed us to output a partition of voters
dethroning~$c$, then output ``control impossible'' and halt.
\end{enumerate}

The two algorithms differ only in the second phase. To describe one
loop iteration for some pair of candidates, $a$ and~$b$, we use the
following notation: For each voter in~$V$, we focus just on his/her
approval of~$a$, $b$, and~$c$, represented (in that order)
as a vector from $\{0,1\}^3$.  Denote 
the number of voters with preference 
$001$ by~$W_c$, with $110$ by~$L_c$, with $100$ by~$S_a$, with $010$
by~$S_b$, with $101$ by~$S_{ac}$, and with $011$ by~$S_{bc}$.  (Voters
with preference $000$ or $111$ 
need not be considered, since they do not affect the difference of Yes
votes among~$a$, $b$, and~$c$.)

\medskip

\noindent
{\bf Loop in model~TE:} For each $a, b \in C$ with
$||\{a,b,c\}||=3$, we test whether we can make $a$ tie or beat $c$ in
$\parpair{C,V_1}$ and make $b$ tie or beat $c$ in~$\parpair{C,V_2}$.

If $W_c - L_c > S_a + S_b$, then this $a$ and $b$
are hopeless, so move on to consider the next $a$ and $b$ in the loop.
Otherwise, we have
\begin{eqnarray}
\label{equ:vulnerable-destructive-voterpartition-approval-te-1}
W_c - L_c & \leq & S_a + S_b .
\end{eqnarray}

Output $\parpair{V_1,V_2}$ as a successful partition and halt, where
$V_1$ contains all voters contributing to $S_{ac}$ and~$S_a$, and also
$\min(W_c, S_a)$ voters contributing to~$W_c$, and where $V_2
= V - V_1$.

In $\parpair{C,V_1}$, $a$ ties or beats~$c$, since $a$ gets 
\[
S_a - \min(W_c, S_a) \geq 0 
\]
more Yes votes than~$c$.  And in
$\parpair{C,V_2}$, $b$ ties or beats~$c$, since $b$ receives 
\[
S_b + L_c - (W_c - \min(W_c, S_a))
\]
more Yes votes than~$c$.  So, for the construction to work, we must
argue that
\[
S_b + L_c + \min(W_c, S_a) - W_c \geq 0.
\]
That is, we need 
\begin{eqnarray}
\label{equ:vulnerable-destructive-voterpartition-approval-te-2}
W_c - L_c & \leq & \min(W_c, S_a) + S_b.
\end{eqnarray}
If $W_c < S_a$,
Equation~(\ref{equ:vulnerable-destructive-voterpartition-approval-te-2})
follows trivially from the fact that $0 \leq L_c + S_b$.  And
if $W_c \geq S_a$,
Equation~(\ref{equ:vulnerable-destructive-voterpartition-approval-te-2})
follows immediately from
Equation~(\ref{equ:vulnerable-destructive-voterpartition-approval-te-1}).

\medskip

\noindent
{\bf Loop in model~TP:} For each $a, b \in C$ with
$||\{a,b,c\}||=3$, 
we test whether we can make $a$ strictly beat $c$ in
$\parpair{C,V_1}$ and make $b$ strictly beat $c$ in $\parpair{C,V_2}$.  

If $W_c - L_c > S_a + S_b - 2$ or $S_a = 0$
or $S_b = 0$, then this $a$ and $b$ are hopeless, so move on
to consider the next $a$ and $b$ in the loop.  Otherwise, we have
\begin{eqnarray}
\label{equ:vulnerable-destructive-voterpartition-approval-tp-1}
W_c - L_c & \leq & S_a + S_b - 2
\end{eqnarray}
and $S_a > 0$ and $S_b > 0$, and output
$\parpair{V_1,V_2}$ as a successful partition and halt, where $V_1$
contains all voters contributing to $S_{ac}$ and~$S_a$, and also
$\min(W_c, S_a - 1)$ voters contributing to~$W_c$, and where
$V_2 = V - V_1$.

In $\parpair{C,V_1}$, $a$ (strictly) beats~$c$, since $a$ gets 
\[
S_a - \min(W_c, S_a - 1) > 0 
\]
more Yes votes than~$c$.  And in $\parpair{C,V_2}$, $b$ (strictly)
beats~$c$, since $b$ has
\[
S_b + L_c - (W_c - \min(W_c, S_a - 1))
\]
more Yes votes than~$c$.  So, for the
construction to work, we must argue that
\[
S_b + L_c + \min(W_c, S_a - 1) - W_c > 0.
\]
That is, we need 
\begin{eqnarray}
\label{equ:vulnerable-destructive-voterpartition-approval-tp-2}
W_c - L_c & < & \min(W_c, S_a - 1) + S_b.
\end{eqnarray}
If $W_c \leq  S_a - 1$,
Equation~(\ref{equ:vulnerable-destructive-voterpartition-approval-tp-2})
reduces to $0 < L_c  + S_b$, which follows from the fact that
in the current case $S_b > 0$.  And if $W_c > S_a -
1$, 
Equation~(\ref{equ:vulnerable-destructive-voterpartition-approval-tp-2})
follows immediately from
Equation~(\ref{equ:vulnerable-destructive-voterpartition-approval-tp-1}).~\end{proofs}

\subsubsection{Adding and Deleting Voters, Destructive Case}

We now turn to proving the vulnerability results for destructive
control by adding and by deleting voters for each of plurality,
Condorcet, and approval voting.  We start with plurality voting.

\begin{theorem}
\label{thm:vulnerable-destructive-voteradditiondeletion-plurality}
Plurality voting is vulnerable/certifiably-vulnerable to destructive
control both by adding voters and by deleting voters.
\end{theorem}

\begin{proofs}
  By
  Examples~\ref{exa:8701:constructive-destructive-voterpartition-plurality-not-immune}
  and~\ref{exa:8704:destructive-voteradditiondeletion-pluralityCondorcet-not-immune},
  susceptibility holds.

In a nutshell, for the adding voters case, we give a ``smart greedy''
algorithm, and for the deleting voters case, we give a ``dumb greedy''
algorithm.  In both cases, we prove only that plurality voting is
certifiably-vulnerable to destructive control, since this implies
vulnerability.  Recall that no ``$k$'' is specified in the corresponding
control problems, as in this setting the chair seeks to determine in
polynomial time the smallest number of voters needed to be added or
deleted to execute control.

In the adding voters case, we are given a set $C$ of candidates, a
distinguished candidate~$c$, a set $V$ of registered voters, and an
additional set $W$ of as yet unregistered voters (both $V$ and $W$ have
preferences over~$C$).  If $c$ already is not a unique plurality
winner in the election~$\parpair{C,V}$, adding no voters accomplishes
our goal, and we are done.  Otherwise, sort all candidates in $C$
distinct from $c$ by how many votes each needs to
tie~$c$.
Let $d_i$ denote the $i$th candidate in the ordering thus obtained, 
and let $\diff{d_i}$ denote $d_i$'s deficit of first-place votes needed
to tie~$c$.
Thus, the order is such that $\diff{d_1} \leq \diff{d_2}
\leq \cdots \leq \diff{d_{||C||-1}}$.  
For $i = 1, 2, \ldots , ||C||-1$, if the number of unregistered voters
whose first choice is $d_i$ is greater than or equal to $\diff{d_i}$, then
add $\diff{d_i}$ of these unregistered voters to ensure that $d_i$ ties $c$
(and $c$ thus is not the unique winner) and halt.  If in no iteration of this
for-loop was some candidate able to dethrone $c$, 
output ``control impossible'' and halt.

In the deleting voters case, we are given a set $C$ of candidates, a
distinguished candidate~$c$, and a set $V$ of voters with preferences
over~$C$.  If $C = \{c\}$, then output ``control impossible'' and
halt; else if $c$ already is not the unique plurality winner in the
election~$\parpair{C,V}$, deleting no voters accomplishes our goal,
and we are done.  Now, if every candidate other than $c$ gets zero
first-place votes, then output ``control impossible'' and halt.
Otherwise, let $d$ be the candidate closest to $c$ in first-place 
votes, and let $\diff{d}$ denote $d$'s deficit of first-place votes
needed to tie~$c$.
Then deleting $\diff{d}$ voters whose first choice is $c$ assures
that $c$ is not the unique winner, and 
this is the fewest deletions that can achieve that.~\end{proofs}

\begin{theorem}
\label{thm:vulnerable-destructive-voteradditiondeletion-Condorcet}
Condorcet voting is vulnerable/certifiably-vulnerable to destructive
control both by adding voters and by deleting voters.
\end{theorem}

\begin{proofs}
  By
  Examples~\ref{exa:8702:destructive-voterpartion-Condorcet-not-immune}
  and~\ref{exa:8704:destructive-voteradditiondeletion-pluralityCondorcet-not-immune},
  susceptibility holds.

We again prove only certifiable vulnerability, since this here implies
vulnerability.  

In the adding voters case, we are given a set $C$ of candidates, a
distinguished candidate~$c$, a set $V$ of registered voters, and an
additional set $W$ of as yet unregistered voters (both $V$ and $W$ have
preferences over~$C$).  If $C = \{c\}$, then output ``control
impossible'' and halt; else if $c$ already is not a Condorcet winner
in the election~$\parpair{C,V}$, adding no candidates accomplishes our
goal, and we are done.  Otherwise, for each candidate $i \neq c$, call
$i$ {\em lucky\/} if and only if the surplus of $c$ relative to $i$
(denoted by $\surplus{c,i}$, which is defined as the number of
registered voters who prefer $c$ to $i$ minus the number of registered
voters who prefer $i$ to~$c$) is less than or equal to the number of
unregistered voters who prefer $i$ to~$c$.  If there is at least one
lucky candidate, then let $d$ be a lucky candidate such that the
surplus of $c$ relative to $d$ is minimum, and add $\surplus{c,d}$
unregistered voters who prefer $d$ to~$c$.  If there exists no lucky
candidate, output ``control impossible'' and halt.

In the deleting voters case, we are given a set $C$ of candidates, a
distinguished candidate~$c$, and a set $V$ of voters with preferences
over~$C$.  If $C = \{c\}$, then output ``control impossible'' and
halt; else if $c$ already is not a Condorcet winner in the
election~$\parpair{C,V}$, deleting no candidates accomplishes our
goal, and we are done.  Otherwise, find a candidate $d$ who comes
closest to $c$ (i.e., relative to whom the surplus of $c$ is minimum),
and delete $\surplus{c,d}$ voters from $V$ who prefer $c$ to~$d$.
Now $c$ and $d$ tie, so $c$ is dethroned.~\end{proofs}

\begin{theorem}
\label{thm:vulnerable-destructive-voteradditiondeletion-approval}
Approval voting is vulnerable/certifiably-vulnerable to destructive
control both by adding voters and by deleting voters.
\end{theorem}

\begin{proofs}
  That susceptibility holds in this case is shown by
  Examples~\ref{exa:8703:destructive-voterpartion-TE-TP-approval-not-immune}
  and~\ref{exa:8705:destructive-voteradditiondeletion-approval-not-immune}.

As before, we prove only certifiable vulnerability, since this here
implies vulnerability.

In the adding voters case, we are given a set $C$ of candidates, a
distinguished candidate~$c$, a set $V$ of registered voters, and an
additional set $W$ of as yet unregistered voters (both $V$ and $W$ have
preferences over~$C$).  If $C = \{c\}$, then output ``control
impossible'' and halt.  Otherwise, if $c$ already is not the unique approval
winner in the election~$\parpair{C,V}$, adding no candidates
accomplishes our goal, and we are done.  Otherwise, for each candidate
$i \neq c$, again define $\surplus{c,i}$ to be the number of Yes votes for
$c$ in $V$ minus the number of Yes votes for $i$ in~$V$.  Among all
candidates $j$ other than $c$ (if any) such that there exist at least
$\surplus{c,j}$ voters in $W$ who vote Yes for $j$ and No for~$c$, let
$d$ be any such $j$ for which $\surplus{c,j}$ is minimum, and add
$\surplus{c,d}$ unregistered voters who vote Yes for $d$ and No
for~$c$.  If no $j$ satisfying the above conditions exists, then
output ``control impossible'' and halt.

In the deleting voters case, we are given a set $C$ of candidates, a
distinguished candidate~$c$, and a set $V$ of voters with preferences
over~$C$.  If $C = \{c\}$, then output ``control impossible'' and
halt.  Otherwise, if $c$ already is not the unique approval winner in the
election~$\parpair{C,V}$, deleting no candidates accomplishes our
goal, and we are done.  
Otherwise, 
let $d$ be a candidate among $C-\{c\}$ for whom $\surplus{c,d}$ is
minimum, and delete $\surplus{c,d}$ voters from $V$ who vote Yes for
$c$ and No for~$d$ (such voters must exist, as they are what is
causing the surplus in the first place).~\end{proofs}

\subsubsection{Adding Candidates, Destructive Case, Condorcet and Approval
  Voting}

Next, we prove that both Condorcet and approval voting are
certifiably-vulnerable (and thus vulnerable) to destructive control by
adding candidates.

\begin{theorem}
\label{thm:vulnerable-destructive-addingcandidates-Condorcetapproval}
Both Condorcet and approval voting are
vulnerable/certifiably-vulnerable to destructive control by adding
candidates.
\end{theorem}

\begin{proofs}
  That susceptibility holds in this case is shown by
  Theorem~\ref{thm:8706:hinged-absolute-corollaries}.

We again prove only certifiable vulnerability, since this here implies
vulnerability.  We are given a set $C$ of qualified candidates and a
distinguished candidate $c \in C$, a set $D$ of possible spoiler
candidates, and a set $V$ of voters with preferences (in the approval
case, the ``preferences'' are 0-1 vectors) over $C \cup D$.

For Condorcet voting, if $c$ already is not the Condorcet winner,
adding no candidates accomplishes our goal, and we are done.
Otherwise, if any spoiler candidate ties or beats~$c$, add one such
candidate and halt.  Otherwise, output ``control impossible'' and halt.

For approval voting,
if $c$ already is not the unique approval
winner in the election $\parpair{C,V}$, adding no candidates
accomplishes our goal, and we are done.  Otherwise, if there exists a
spoiler candidate $d$ who ties or beats $c$ among the voters in $V$ in
Yes votes, add one such spoiler candidate and halt.  Otherwise, output
``control impossible'' and halt.~\end{proofs}

\subsubsection{Deleting Candidates, Partition and Run-off Partition of
  Candidates, Constructive Case, Approval Voting}

Finally, we show the vulnerability results for approval voting
for constructive control by
deleting candidates, and by partition of candidates and run-off partition of
candidates, both in model~TE\@.  

\begin{theorem}
\label{thm:vulnerable-constructive-deletingpartitioncandidates-approval}
Approval voting is vulnerable/certifiably-vulnerable to constructive
control by deleting candidates, partition of candidates in model~TE,
and run-off partition of candidates in model~TE\@.
\end{theorem}

\begin{proofs}
  That susceptibility holds in this case is shown by
  Theorem~\ref{thm:8706:hinged-absolute-corollaries}
  and 
  Example~\ref{exa:8707:constructive-deletingpartitioncandidates-approval-not-immune}.

As in the previous proofs, we only show certifiable vulnerability,
which again implies vulnerability.  Thus, no ``$k$'' is specified in
the control problem corresponding to the deleting candidates case, and
in all three cases we are given a set $C$ of candidates, a
distinguished candidate~$c$, and a set $V$ of registered voters.  We
now describe a polynomial-time algorithm for each of the three
constructive control problems considered.

In the deleting candidates case, 
if $c$ already is 
the unique approval winner in the election~$\parpair{C,V}$, deleting
no candidates accomplishes our goal, and we are done.  Otherwise,
delete every candidate other than $c$ 
who has at least as many Yes votes as $c$ has 
in $V$ and halt.

In the partition of candidates case, 
if $c$ already is 
the unique approval winner, then output $\parpair{\emptyset,C}$ as a
successful partition and halt.  Otherwise, for each candidate $a \in C$, let
$y_a$ denote the number of Yes votes cast for $a$ in~$V$, and let $Y =
\max\{ y_a \condition a \in C\}$.

Now, if there exists \emph{exactly} one $a \in C - \{c\}$
such that $y_a = Y$, then
output ``control impossible'' and halt, since $c$ cannot be made the
unique winner in this case.  On the other hand, if there
exist at least two distinct candidates in $C - \{c\}$ whose
number of Yes votes is~$Y$, then output $\parpair{C_1,C_2}$ with $C_1
= C - \{c\}$ and $C_2 = \{c\}$ as a successful partition and halt.
This works, since in subelection $\parpair{C_1,V}$ all
candidates are eliminated.

Note that the same algorithm also works for the
run-off partition of candidates case in model~TE\@.~\end{proofs}

\subsection{Proving Resistance}
\label{sec:resistance-proofs}

The resistance results are based on clear containments in NP,
plus (polynomial-time many-one) reductions establishing NP-hardness.

The following lemma says that
for the voting systems considered here (though that may be different in
general), whenever the corresponding control problem is
$\np$-hard, immunity cannot hold unless $\p = \np$.

\begin{lemma}
\label{lem:nonimmunity-not-needed-for-resistance-proofs}
For each voting system for which winnership can be tested in
polynomial time, if the control problem corresponding to one of the
settings considered here is $\np$-hard, then the system cannot be immune
to control in this setting unless $\p = \np$. 
\end{lemma}

\begin{proofs}
Consider any voting system for which winner-testing (``Is $c$ a winner?'')
can be done in
polynomial time.  Suppose that the decision problem associated with
any one of the control scenarios defined in Section~\ref{sec:results}
is $\np$-hard.  Then, as mentioned in Footnote~\ref{foo:longfootnote},
if immunity were to hold, the associated decision problem would be
in~$\p$, which would imply $\p = \np$.~\end{proofs}

However, proving immunity and susceptibility under assumptions
regarding $\p$-versus-$\np$ is obviously less attractive than 
proving immunity and susceptibility
unconditionally.
In particular, the ideal first
step toward proving resistance results is to prove, via theorems or
examples, susceptibility to the corresponding types of control.  We
have done that in Section~\ref{sec:immune-and-others}, and will invoke
items from that section here.

\subsubsection{Plurality  Voting}

We whenever possible try to achieve multiple resistance results via a
single proof.
For example,
with a single proof we establish the key part of 
all seven resistance results for
plurality voting: destructive
control by adding,
deleting, partition (TE and TP), and 
run-off partition  (TE and TP) of candidates,\footnote{Our constructions
ensure that the distinguished candidate is never tied for winner
in any subelection 
in the image of the $\np$-hardness reduction.
Thus, these results hold both in the Ties-Eliminate and
Ties-Promote models.}
and by partition of voters (TP).
We now provide this proof, which is achieved via one general construction that
yields the reductions, each from the $\np$-complete problem Hitting
Set, see Garey and Johnson~\cite{gar-joh:b:int}.

\subsubsection*{Hitting Set}

\begin{description}
\item[Given:] A set $B = \{b_1, b_2, \ldots , b_m\}$, a family
$\mathcal{S} = \{S_1, S_2, \ldots , S_n\}$ of subsets $S_i$ of~$B$,
and a positive integer~$k$.
\item[Question:] Does $\mathcal{S}$ have a hitting set of size at
most~$k$?  That is, is there a set $B' \subseteq B$ with $||B'|| \leq
k$ such that for each~$i$, $S_i \cap B' \neq \emptyset$?
\end{description}

We now present our general construction for the destructive control
problems related to plurality voting.

\begin{construction}[Construction of an Election from a Hitting Set Instance]
\label{con:general-construction-resistance-plurality-destructive}
\quad
Given a triple $\parpair{B,\mathcal{S},k}$, where $B = \{b_1, b_2,
\ldots , b_m\}$ is a set, $\mathcal{S} = \{S_1, S_2, \ldots , S_n\}$
is a family of subsets $S_i$ of~$B$, and $k \leq m$
is a positive integer, we
construct the following election:
\begin{itemize}
\item The candidate set is $C = B \cup \{c,w\}$.
\item The voter set $V$ is defined as follows:
\begin{itemize}
\item There are $2(m-k) + 2n(k+1) + 4$ voters of the form $c > w >
\cdots$, where ``$\cdots$'' means that the remaining candidates follow
in some arbitrary order.
\item There are $2n(k+1) + 5$ voters of the form $w > c > \cdots$.
\item For each~$i$, $1 \leq i \leq n$, there are $2(k+1)$ voters of
the form $S_i > c > \cdots$, where ``$S_i$'' denotes the elements of
$S_i$ in some arbitrary order.
\item Finally, for each~$j$, $1 \leq j \leq m$, there are two voters
of the form $b_j > w > \cdots$.
\end{itemize}
\end{itemize}
\end{construction}

We now show that the election $\parpair{C,V}$ constructed above has
some useful properties needed to establish resistance to destructive
control for plurality voting in the seven settings mentioned.
For every candidate~$d$, let $\score{d}$ denote
the number of voters who rank $d$ first in a given election.

\begin{claim}
\label{cla:53-A}
If $B'$ is a hitting set of $\mathcal{S}$ of size $k$,
then $w$ is the unique plurality winner of the election
$\parpair{B' \cup \{c,w\},V}$.
\end{claim}

\begin{proofs}
If $B'$ is a hitting set of $\mathcal{S}$ of size~$k$, then in the
election $\parpair{B' \cup \{c,w\},V}$, we have
\begin{eqnarray*}
\score{c}   & =    & 2(m-k) + 2n(k+1) + 4,  \\
\score{w}   & =    & 2n(k+1) + 5 + 2(m-k),\quad \mbox{and} \\
\score{b_j} & \leq & 2n(k+1) + 2          \quad \mbox{for each~$j$.}
\end{eqnarray*}
It follows that $w$ is the unique plurality winner of the election
$\parpair{B' \cup \{c,w\},V}$.~\end{proofs}

\begin{claim}
\label{cla:53-B}
Let $D \subseteq B \cup \{w\}$.
If $c$ is not the unique plurality winner of election $\parpair{D
\cup \{c\},V}$, then there exists a set $B' \subseteq B$ such that
\begin{enumerate}
\item $D = B' \cup \{w\}$, 
\item $w$ is the unique plurality winner of the election $\parpair{B' \cup
    \{c,w\},V}$, and
\item $B'$ is a hitting set of $\mathcal{S}$ of size less than or
  equal to~$k$.
\end{enumerate}
\end{claim}

\begin{proofs}
Let $D \subseteq B \cup \{w\}$ and
  suppose that $c$ is not the unique plurality winner of election
  $\parpair{D \cup \{c\},V}$. We show the
  three properties stated in the claim.
  
  First note that for all $b \in D
  \cap B$, $\score{b} < \score{c}$ in $\parpair{D \cup \{c\},V}$.  Since
  $c$ is not the unique plurality winner of $\parpair{D \cup \{c\},V}$, it
  follows that
  $w \in D$ and $\score{w} \geq \score{c}$.  Let $B' \subseteq B$ be such that
  $D = B' \cup \{w\}$.   Then $D \cup \{c\} = B' \cup \{c,w\}$.
Since $\score{w}$ is odd and $\score{c}$ is even, it
  follows that $w$ is the unique plurality winner of $\parpair{B' \cup
    \{c,w\},V}$.  This proves the first two properties stated.
  
  To prove the third property, note that in $\parpair{B' \cup \{c,w\},V}$, we
  have
\begin{eqnarray*}
\score{w} & = & 2n(k+1) + 5 + 2(m - ||B'||) \quad \mbox{and} \\
\score{c} & = & 2(m-k) + 2n(k+1) + 4 + 2(k+1)\ell,
\end{eqnarray*}
where $\ell$ is the number of sets in $\mathcal{S}$ that are not hit
by $B'$ (i.e., that have an empty intersection with~$B'$).  Since
$\score{c} \leq \score{w}$, it follows that
\[
2(m-k) + 2(k+1)\ell \leq 1 + 2(m - ||B'||),
\]
which implies $(k+1)\ell + ||B'|| - k \leq 0$. So $\ell = 0$.  Thus,
$B'$ is a hitting set of $\mathcal{S}$ of size at most~$k$, which
proves the third property.~\end{proofs}

Next, we show that
Construction~\ref{con:general-construction-resistance-plurality-destructive}
yields a polynomial-time many-one reduction from Hitting Set to Destructive
Control by Adding Candidates for plurality voting.

\begin{claim}
\label{cla:54-A}
$\mathcal{S}$ has a hitting set of size less than or equal to $k$ if and only
if destructive control by adding candidates can be executed for the election
with qualified candidates $\{c,w\}$, spoiler candidates~$B$, distinguished
candidate~$c$, and voter set~$V$.
\end{claim}

\begin{proofs}
If $\mathcal{S}$ has a hitting set of size less than or equal to $k$, 
then
since $k \leq m$, $\mathcal{S}$ has a hitting set of size $k$. 
Thus, the implication from left to right follows from Claim~\ref{cla:53-A}.
The implication from right to left follows from Claim~\ref{cla:53-B}.~\end{proofs}

So from this and 
Theorem~\ref{thm:8706:hinged-absolute-corollaries}
we have the following.

\begin{corollary}
\label{cor:resistance-destructive-addingcandidates-plurality}
Plurality voting is resistant to destructive control by adding candidates.
\end{corollary}

By a similar argument, Hitting Set can be reduced to Destructive Control by
Deleting Candidates for plurality voting.

\begin{claim}
\label{cla:54-B}
$\mathcal{S}$ has a hitting set of size at most $k$ if and only if the
election with candidate set~$C$, distinguished candidate~$c$, and voter set
$V$ can be destructively controlled by deleting at most $m-k$ candidates.
\end{claim}

\begin{proofs}
  Let $B'$ be a hitting set of $\mathcal{S}$ of size
  $k$.
 By Claim~\ref{cla:53-A}, $c$ is not the unique plurality winner of
  the election $\parpair{B' \cup \{c,w\},V}$.  Since $B' \cup \{c,w\} = C - (B
  - B')$, $||B|| = m$, and $||B'|| = k$, the right-hand side of the
  equivalence follows.
  
  For the converse, let $D \subseteq B \cup \{w\}$ be such that $||D|| \leq
  m-k$, and suppose that $c$ is not the unique plurality winner of
  $\parpair{C-D,V}$. Since $c \in C-D$, it follows from
Claim~\ref{cla:53-B} that
  $(C - D) - \{c\} = B' \cup \{w\}$, where $B'$
  is a hitting set of $\mathcal{S}$ of size less than or equal to~$k$.~\end{proofs}

So from this and 
Example~\ref{exa:8708:destructive-deletingpartitioncandidates-plurality-not-immune}
we have the following.

\begin{corollary}
\label{cor:resistance-destructive-deletingcandidates-plurality}
Plurality voting is resistant to destructive control by deleting candidates.
\end{corollary}

Now, we show that
Construction~\ref{con:general-construction-resistance-plurality-destructive}
also yields a polynomial-time many-one reduction from Hitting Set to
Destructive Control by Partition of Candidates for plurality voting.

\begin{claim}
\label{cla:56-B}
$\mathcal{S}$ has a hitting set of size at most $k$ if and only if the
election with candidate set~$C$, distinguished candidate~$c$, and voter set
$V$ can be destructively controlled by partition of candidates (both in
model~TE and~TP).
\end{claim}

\begin{proofs}
  Let $B'$ be a hitting set of $\mathcal{S}$ of size~$k$.
  Partition $C$ into $C_1 = B' \cup \{c,w\}$ and $C_2 = B - B'$.  By
  Claim~\ref{cla:53-A}, $w$ is the unique plurality winner of
  $\parpair{C_1,V}$, and $c$ thus cannot win the election $\parpair{C,V}$.
  
  For the converse, suppose that there exists a partition of candidates
  such that $c$ is not the unique plurality winner of the two-stage
  election corresponding to that partition.  Then, certainly,
  there exists a set $D \subseteq B \cup \{w\}$
  such that $c$ is not the unique plurality winner of $\parpair{D \cup
  \{c\},V}$.  By Claim~\ref{cla:53-B}, $\mathcal{S}$ has a hitting set of
  size at most~$k$.~\end{proofs}

So from this and 
Example~\ref{exa:8708:destructive-deletingpartitioncandidates-plurality-not-immune}
we have the following.

\begin{corollary}
\label{cor:resistance-destructive-partitioncandidates-plurality}
Plurality voting is resistant to destructive control by partition of
candidates (both in model~TE and~TP).
\end{corollary}

The same argument works for proving that plurality voting is resistant to
destructive control by run-off partition of candidates, again by a reduction
from Hitting Set.

\begin{claim}
\label{cla:55-B}
$\mathcal{S}$ has a hitting set of size at most $k$ if and only if the
election with candidate set~$C$, distinguished candidate~$c$, and voter set
$V$ can be destructively controlled by run-off partition of candidates
(both in model~TE and~TP).
\end{claim}

\begin{proofs}
  Let $B'$ be a hitting set of $\mathcal{S}$ of size~$k$.  Partition
  $C$ into $C_1 = B' \cup \{c,w\}$ and $C_2 = B - B'$.  By
  Claim~\ref{cla:53-A}, $w$ is the unique plurality winner of
  $\parpair{C_1,V}$, and $c$ thus cannot win the election $\parpair{C,V}$.

  For the converse, suppose that there exists a partition of candidates
  such that $c$ is not the unique plurality winner in the run-off
  election corresponding to that partition.  Then, certainly,
  there exists a set $D \subseteq B \cup \{w\}$
  such that $c$ is not the unique plurality winner of $\parpair{D \cup
  \{c\},V}$.  By Claim~\ref{cla:53-B}, $\mathcal{S}$ has a hitting set of
  size at most~$k$.~\end{proofs}

So from this and 
Example~\ref{exa:8708:destructive-deletingpartitioncandidates-plurality-not-immune}
we have the following.

\begin{corollary}
\label{cor:resistance-destructive-run-off-partitioncandidates-plurality}
Plurality voting is resistant to destructive control by run-off partition of
candidates (both in model~TE and~TP).
\end{corollary}

Finally, we show that plurality voting is resistant to both constructive and
destructive control by
partition of voters in the TP model.  To this end, we reduce from the Hitting
Set problem restricted to instances where $n(k+1) + 1 \leq m-k$.  We first
define this restriction and prove that it still is $\np$-complete.

\subsubsection*{Restricted Hitting Set}

\begin{description}
\item[Given:] A set $B = \{b_1, b_2, \ldots , b_m\}$, a family
$\mathcal{S} = \{S_1, S_2, \ldots , S_n\}$ of subsets $S_i$ of~$B$,
and a positive integer~$k$ such that $n(k+1) + 1 \leq m-k$.
\item[Question:] Does $\mathcal{S}$ have a hitting set of size at
most~$k$?  That is, is there a set $B' \subseteq B$ with $||B'|| \leq
k$ such that for each~$i$, $S_i \cap B' \neq \emptyset$?
\end{description}

\begin{theorem}
\label{thm:restricted-hitting-set-is-np-complete}
 Restricted Hitting Set is $\np$-complete.
\end{theorem}

\begin{proofs}
  Restricted Hitting Set clearly is in~$\np$.  To show that it is $\np$-hard,
  we reduce Hitting Set to Restricted Hitting Set.  Let
  $\parpair{\widehat{B},\widehat{\mathcal{S}},k}$ be a Hitting Set instance,
  where
\begin{eqnarray*}
\widehat{B}           & = & \{b_1, b_2, \ldots , b_{\widehat{m}}\}, \\
\widehat{\mathcal{S}} & = & \{\widehat{S}_1, \widehat{S}_2, \ldots , 
                              \widehat{S}_n\},
\end{eqnarray*}
$\widehat{S}_i \subseteq \widehat{B}$ for each~$i$, $1 \leq i \leq n$,
and $k + 1 \leq \widehat{m}$.  Define
an instance of Restricted Hitting Set $\parpair{B,\mathcal{S},k}$, where
\begin{eqnarray*}
B           & = & \widehat{B} \cup \{a_{i,j} \condition 
                  1 \leq i \leq n \mbox{ and } 1 \leq j \leq k+1\}, \\
S_i         & = & \widehat{S}_i \cup \{a_{i,1}, a_{i,2}, \ldots , a_{i,k+1}\},
                  \quad \mbox{for $1 \leq i \leq n$, and } \\
\mathcal{S} & = & \{S_1, S_2, \ldots , S_n\} .
\end{eqnarray*}
It is immediate that $\widehat{\mathcal{S}}$ has a
hitting set of size $k$ if and only if $\mathcal{S}$ has a hitting set of
size~$k$.

Let $m = ||B|| = \widehat{m} + n(k+1)$.  Since $k+1 \leq \widehat{m}$,
we have
\[
n(k+1) + k+1 \leq n(k+1) + \widehat{m} = m,
\]
i.e., $n(k+1) + 1 \leq m-k$.~\end{proofs}

\begin{claim}
\label{cla:62-A}
In the election $\parpair{C,V}$ from
Construction~\ref{con:general-construction-resistance-plurality-destructive},
if $n(k+1) + 1 \leq m-k$ then for every partition of $V$ into $V_1$ and~$V_2$,
$c$ is a plurality winner of $\parpair{C,V_1}$ or of~$\parpair{C,V_2}$.
\end{claim}

\begin{proofs}
  For a contradiction, suppose that $c$ is a winner of neither
  $\parpair{C,V_1}$ nor~$\parpair{C,V_2}$.  For each $U \seq V$ and for each
  $i \in C$, let $\scoresub{U}{i}$ denote the number of first-place votes that
  $i$ has in $\parpair{C,U}$.
  Let $x \in B \cup \{w\}$ be a
  winner of $\parpair{C,V_1}$, and let $y \in B
  \cup \{w\}$ be a winner of $\parpair{C,V_2}$.
  Then
\begin{eqnarray}
\label{equ:62-A}
\scoresub{V_1}{x} + \scoresub{V_2}{y} \geq \scoresub{V}{c} + 2.
\end{eqnarray}
Since $c$'s score in $(C,V)$ is greater than that of any other candidate, we
have $x \neq y$.  It follows that
\begin{eqnarray*}
\scoresub{V_1}{x} + \scoresub{V_2}{y}
 & \leq & \scoresub{V}{w} + \scoresub{V}{b_i} \\
 & \leq & 2n(k+1) + 5 + 2n(k+1) + 2 \\
 & \leq & 2n(k+1) + 5 + 2(m-k) \\
 & =    & \scoresub{V}{c} + 1,
\end{eqnarray*}
which contradicts Equation~(\ref{equ:62-A}).  Thus, $c$ is a winner
of $\parpair{C,V_1}$ or of $\parpair{C,V_2}$.~\end{proofs}

We now show that
Construction~\ref{con:general-construction-resistance-plurality-destructive}
also provides a reduction from Restricted Hitting Set both to Constructive
Control by Partition of Voters and to Destructive Control by Partition of
Voters in the Ties-Promote model for plurality voting.

\begin{claim}
\label{cla:63-A:64-A}
In the election $\parpair{C,V}$ from
Construction~\ref{con:general-construction-resistance-plurality-destructive},
if $n(k+1) + 1 \leq m-k$ then
the following three statements are equivalent:
\begin{enumerate}
\item $\mathcal{S}$ has a hitting set of size at most $k$.
\item $V$ can be partitioned such that $w$ is the unique plurality winner in
  the TP model.
\item $V$ can be partitioned such that $c$ is not the unique plurality winner
  in the TP model.
\end{enumerate}
\end{claim}

\begin{proofs}
  To show that the first statement implies the second statement, let $B'$ be a
  hitting set of $\mathcal{S}$ of size~$k$.  Partition $V$ into $V_1$
  and~$V_2$, where $V_1$ consists of one voter of the form $w > c >
  \cdots$ and for every $b \in B'$ one voter of the form
  $b > w > \cdots$,
and where $V_2 = V - V_1$.  Then the candidates in $B' \cup \{w\}$ are the
winners of  $\parpair{C,V_1}$ and move forward to the run-off in
  the TP model, and $c$ is the winner of $\parpair{C,V_2}$.  
By Claim~\ref{cla:53-A}, $w$ is the unique plurality
  winner of the final election $\parpair{B' \cup \{c,w\},V}$.  
  
  Clearly, if $w$ is the unique plurality winner for some partition of $V$ in
  the TP model, then $c$ cannot be the unique plurality winner of this
  election for the same partition.  Thus, the second statement implies the
  third statement.
  
  Finally, we show that the third statement implies the first statement.
  Suppose there is a partition of $V$ such that $c$ is not the unique
  plurality winner of the election in the TP model.  By
  Claim~\ref{cla:62-A}, $c$ is a winner of one of the subelections and will
  thus participate in the final run-off.
  It follows that $c$ is not the unique winner of a run-off election
  involving~$c$, i.e., $c$ is not the unique winner in
  $\parpair{D \cup \{c\},V}$, for some $D \subseteq B \cup \{w\}$.  By
  Claim~\ref{cla:53-B}, $\mathcal{S}$ has a hitting set of size at most~$k$.
  This completes the proof.~\end{proofs}

Theorem~\ref{thm:restricted-hitting-set-is-np-complete},
Claim~\ref{cla:63-A:64-A}, and
Example~\ref{exa:8709:constructive-destructive-voterpartition-TP-plurality-not-immune}
have the following corollary.

\begin{corollary}
\label{cor:resistance-constructivedestructive--partitionvoters-TP-plurality}
\begin{enumerate}
\item Plurality voting is resistant to constructive control by partition of
  voters in model~TP\@.
\item Plurality voting is resistant to destructive control by partition of
  voters in model~TP\@.
\end{enumerate}
\end{corollary}

\subsubsection{Approval  Voting, Constructive Case, Voter Control}

For approval voting, our reductions proving resistance are from the
$\np$-complete problem Exact Cover by Three-Sets (X3C, for short), see Garey
and Johnson~\cite{gar-joh:b:int}.

\subsubsection*{Exact Cover by Three-Sets (X3C)}

\begin{description}
\item[Given:] A set $B = \{b_1, b_2, \ldots , b_m\}$, where $m = 3k$ for a
  positive integer~$k$, and a family $\mathcal{S} = \{S_1, S_2, \ldots ,
  S_n\}$ of subsets $S_i$ of $B$ with $||S_i|| = 3$ for each~$i$.
\item[Question:] Does $\mathcal{S}$ have an exact cover for~$B$?  That is, is
  there a subfamily $\mathcal{S}' \subseteq \mathcal{S}$ such that every
  element of $B$ occurs in exactly one set in~$\mathcal{S}'$?
\end{description}

\begin{theorem}
\label{thm:resistance-constructive-addingvoters-approval}
Approval voting is resistant to constructive control by adding voters.
\end{theorem}

\begin{proofs}
  That susceptibility holds in this case is shown by
  Example~\ref{exa:8703:destructive-voterpartion-TE-TP-approval-not-immune}.

  Given an instance $\parpair{B,\mathcal{S}}$ of~X3C, where $B = \{b_1, b_2,
  \ldots , b_m\}$, $m = 3k$, $k > 1$,
 $\mathcal{S} = \{S_1, S_2, \ldots , S_n\}$, and
  $S_i \subseteq B$ with $||S_i|| = 3$ for each~$i$, $1 \leq i \leq n$,
  construct the following instance of Constructive Control by Adding Voters
  for approval voting:
\begin{itemize}
\item The candidate set is $C = B \cup \{w\}$, where $w$ is the distinguished
  candidate.
\item $V$ consists of $k-2$ registered voters who each approve of $b_1, b_2,
  \ldots , b_m$ and disapprove of~$w$.
\item $W$ consists of $n$ unregistered voters: For each~$i$, $1 \leq i \leq
  n$, there is one voter in $W$ who approves of $w$ and the three candidates
  in~$S_i$, and who disapproves of all other candidates.
\end{itemize}

We claim that $\mathcal{S}$ contains an exact cover for $B$ if and only if $w$
can be made the unique approval winner by adding at most $k$ voters.

For the left to right direction, simply add the $k$ voters from $W$ that
correspond to the exact
cover for~$B$.  Then $w$ has $k$ Yes votes and every $b \in B$ has $(k-2) + 1 =
k-1$ Yes votes, so $w$ is the unique approval winner.

For the right to left direction,
suppose that $w$ can be made the unique approval winner by
adding at most $k$ voters.  Then we clearly need to add exactly $k$ voters
and every $b \in B$ can gain at most one Yes vote.
Since each voter in $W$
casts
three Yes votes for candidates in $B$, it follows that every $b \in B$ gains
exactly one Yes vote. Thus, the $k$ added voters correspond to an
exact cover for~$B$.~\end{proofs}

\begin{theorem}
\label{thm:resistance-constructive-deletingvoters-approval}
Approval voting is resistant to constructive control by deleting voters.
\end{theorem}

\begin{proofs}
  That susceptibility holds in this case is shown by
  Example~\ref{exa:8705:destructive-voteradditiondeletion-approval-not-immune}.

  Let an instance $\parpair{B,\mathcal{S}}$ of X3C be given, where $B = \{b_1,
  b_2, \ldots , b_m\}$, $m = 3k$, $k > 0$, $\mathcal{S} = \{S_1, S_2, \ldots , S_n\}$,
  and $S_i \subseteq B$ with $||S_i|| = 3$ for each~$i$, $1 \leq i \leq n$.
  For each~$j$, $1 \leq j \leq m$, let
\begin{eqnarray*}
\ell_j      & = & ||\{S_i \in \mathcal{S} \condition b_j \in S_i\}||.
\end{eqnarray*}

Construct the following election:
\begin{itemize}
\item The candidate set is $C = B \cup \{w\}$, where $w$ is the distinguished
  candidate.
\item The voter set $V$ consists of the following voters:
\begin{itemize}
\item For each~$i$, $1 \leq i \leq n$, there is one voter in $V$ who
  approves of all candidates in $S_i$ and who disapproves of all other candidates.
\item There are $n$ voters $v_1, v_2, \ldots, v_n$ in $V$
  such that, for each $i$, $1 \leq i \leq n$, $v_i$ approves of~$w$,
  and $v_i$ approves of $b_j$ if and only if $i \leq n - \ell_j$.
\end{itemize}
\end{itemize}

Note that the election $\parpair{C,V}$ has the property that all candidates
have $n$ Yes votes.

We claim that $\mathcal{S}$ contains an exact cover for $B$ if and only if $w$ can
be made the unique approval winner by deleting at most $k$ voters.

For the left to right direction, simply delete the $k$ voters from $V$ that correspond
to an exact cover for~$B$.  Then every $b \in B$ loses one Yes vote, leaving $w$
the unique approval winner.

For the right to left direction,
suppose that $w$ can be made the unique approval
winner by deleting at most $k$ voters.  Without
loss of generality, we may assume that none of the deleted voters
approves of $w$.
So, we assume that only voters corresponding to $S_i$'s have
been deleted.
For $w$ to have become the unique winner, every $b \in B$
must have lost at least one Yes vote.  It follows that the deleted
voters correspond to a cover, and since the cover has size at most $k$,
this must be an exact cover for~$B$.~\end{proofs}

\begin{theorem}
\label{thm:resistance-constructive-partitionvoters-TP-approval}
Approval voting is resistant to constructive control by partition of voters
in model~TP\@.
\end{theorem}

\begin{proofs}
  That susceptibility holds in this case is shown by
  Example~\ref{exa:8710:constructive-addingdeletingvoterpartition-approval-not-immune}.

  Let an instance $\parpair{B,\mathcal{S}}$ of X3C be given, where $B = \{b_1,
  b_2, \ldots , b_m\}$, $m = 3k$, $k > 0$, $\mathcal{S} = \{S_1, S_2, \ldots , S_n\}$,
  and $S_i \subseteq B$ with $||S_i|| = 3$ for each~$i$, $1 \leq i \leq n$.
  We modify the construction from the proof of
  Theorem~\ref{thm:resistance-constructive-deletingvoters-approval}.
  As in that proof, for each~$j$, $1 \leq j \leq m$, let
\begin{eqnarray*}
\ell_j      & = & ||\{S_i \in \mathcal{S} \condition b_j \in S_i\}||.
\end{eqnarray*}

Now, define the following election:
\begin{itemize}
\item The candidate set is $C = B \cup \{w,x,y\}$,
where $w$ is the distinguished
  candidate.
\item The voter set $V$ consists of the following voters:
\begin{itemize}
\item For each~$i$, $1 \leq i \leq n$, there is one voter in $V$ who
  approves of $y$ and of all elements of $S_i$ and who disapproves of all other candidates.
\item There are $n$ voters $v_1, v_2, \ldots ,v_n$ in $V$
  such that, for each $i$, $1 \leq i \leq n$, $v_i$ approves
  of~$w$, $v_i$ disapproves of~$x$, $v_i$ disapproves
of $y$,  and $v_i$ approves of $b_j$ if and only if
  $i \leq n - \ell_j$.
\item There are $k+1$ voters in $V$ who approve of $x$ and disapprove of all
  other candidates.
\item Finally, there are $k+2$ voters in $V$ who disapprove of $x$ and approve
  of all other candidates.
\end{itemize}
\end{itemize}

Note that this election has the property that all candidates
other than $x$ have $n + k + 2$ Yes votes.

We claim that $\mathcal{S}$ contains an exact cover for $B$ if and only if
$w$ can
be made the unique approval winner by partition of voters in model TP\@.

For the left to right direction, if $\mathcal{S}$ contains an exact cover
for~$B$, then let $V_2$ consist of the $k$ voters corresponding to the sets
in the cover and of all the $k+1$ voters who approve of only~$x$,
and let $V_1 = V - V_2$.
Then
\begin{itemize}
\item $w$ is the unique approval winner of $\parpair{C,V_1}$,
\item $x$ is the unique approval winner of $\parpair{C,V_2}$, and
\item $w$ wins the run-off against~$x$.
\end{itemize}

For the right to left direction,
suppose that $w$ can be made the unique approval winner by
partition of voters in model TP\@.
Since $w$ is the unique winner in the run-off, 
and since every candidate other than $x$ is tied with $w$
(each having $n + k+2$ Yes votes in~$V$), the only candidates
that can participate in
the run-off are $w$ and~$x$.
Since we are in the TP model, $w$ must be the unique
winner of one of the subelections and $x$ must
be the unique winner of the other subelection. 

Let $\parpair{V_1,V_2}$ be a partition of~$V$ such that
$w$ is the unique winner of $\parpair{C,V_1}$ and
such that $x$ is the unique winner of $\parpair{C,V_2}$.
As in the
proof of Theorem~\ref{thm:resistance-constructive-deletingvoters-approval}, it
follows that the voters corresponding to $S_i$'s that are not in $V_1$
(i.e., that are in~$V_2$) correspond to a cover. 
Since $x$ is the unique winner of $(C,V_2)$ and 
$x$ has $k+1$ Yes votes, $y$ can have at most $k$ Yes votes in 
$V_2$.  It follows that there are at most $k$ voters corresponding
to $S_i$'s in $V_2$.  Thus, there are exactly $k$ such voters, and
these voters correspond to an exact cover.~\end{proofs}

Note that the previous construction won't work for the TE model,
since in that model, $w$ also wins the election if two or
more candidates are tied for first place in $V_2$.
In the proof
of the next theorem, we will adapt the 
construction from the proof of
Theorem~\ref{thm:resistance-constructive-partitionvoters-TP-approval}.

\begin{theorem}
\label{thm:resistance-constructive-partitionvoters-TE-approval}
Approval voting is resistant to constructive control by partition of voters
in model~TE\@. 
\end{theorem}

\begin{proofs}
  That susceptibility holds in this case is shown by
  Example~\ref{exa:8710:constructive-addingdeletingvoterpartition-approval-not-immune}.

  Let an instance $\parpair{B,\mathcal{S}}$ of X3C be given, where $B = \{b_1,
  b_2, \ldots , b_m\}$, $m = 3k$, $k > 0$, $\mathcal{S} = \{S_1, S_2, \ldots , S_n\}$,
  and $S_i \subseteq B$ with $||S_i|| = 3$ for each~$i$, $1 \leq i \leq n$.
  We modify the construction from the proof of
  Theorem~\ref{thm:resistance-constructive-partitionvoters-TP-approval}.
  As in that proof, for each~$j$, $1 \leq j \leq m$, let
\begin{eqnarray*}
\ell_j      & = & ||\{S_i \in \mathcal{S} \condition b_j \in S_i\}||.
\end{eqnarray*}

Now, define the following election:
\begin{itemize}
\item The candidate set is $C = B \cup \{w,x,y\} \cup \{z_1, \ldots, z_n\}$,
where $w$ is the distinguished
  candidate.
\item The voter set $V$ consists of the following voters:
\begin{itemize}
\item For each~$i$, $1 \leq i \leq n$, there is one voter in $V$ who
  approves of $y$ and of all elements of $S_i$ and who disapproves of all other candidates.
\item For each~$i$, $1 \leq i \leq n$, there is one voter in $V$ who
approves of $y$ and $z_i$ and who disapproves of  all
other candidates.
\item There are $n$ voters $v_1, v_2, \ldots ,v_{n}$ in $V$
  such that, for each $i$, $1 \leq i \leq n$, $v_i$ approves
  of~$w$, $v_i$ disapproves of~$x$, $v_i$ disapproves
of $y$, $v_i$ approves of $b_j$ if and only if
  $i \leq n - \ell_j$,
 and $v_i$ approves of $z_j$ if and only if
$i \neq n$.
\item There are $n + k$ voters in $V$ who approve of $x$ and
who  disapprove of all other candidates.
\end{itemize}
\end{itemize}

Note that this election has the property that all candidates
other than $x$ and $y$ have $n$ Yes votes.

We claim that $\mathcal{S}$ contains an exact cover for $B$ if and only if
$w$ can
be made the unique approval winner by partition of voters in model TE\@.

For the left to right direction, if $\mathcal{S}$ contains an exact cover
for~$B$, then let $V_2$ consist of the $k$ voters corresponding to the sets
in the cover and of all the $n+k$ voters who approve of only~$x$ and
for each $i$, $1 \leq i \leq n$, of the voter who
approves of only $y$ and $z_i$. Let $V_1 = V - V_2$.
Then $w$ is the unique approval winner of $\parpair{C,V_1}$, and
$x$ and $y$ are tied for first place in $\parpair{C,V_2}$
with $n + k$ Yes votes each. Since
we are in model TE, no candidates are nominated by $\parpair{C,V_2}$,
and $w$ wins the run-off (and thus the election) by default.

For the right to left direction,
suppose that $w$ can be made the unique approval winner by
partition of voters in model TE\@.  Since we are in model
TE, $w$ must be the unique winner of one of the subelections.
Let $\parpair{V_1,V_2}$ be a partition of~$V$ such that
$w$ is the unique winner of $\parpair{C,V_1}$.
As in the
proof of Theorem~\ref{thm:resistance-constructive-deletingvoters-approval}, it
follows that the voters corresponding to $S_i$'s that are not in $V_1$
(i.e., that are in~$V_2$) correspond to a cover.

Suppose that there are more than $k$ voters that correspond to $S_i$'s
in $V_2$.  Note 
that for each $i$, $1 \leq i \leq n$, the voter that approves of only
$y$ and $z_i$ must also be in $V_2$ (for if it weren't, $z_i$ would have
at least as many Yes votes in $V_1$ as $w$).  It follows that
$y$ has more than $n + k$ Yes votes in $V_2$.  But then $y$
is the unique approval winner in $V_2$, since no other candidate
has more than $n + k$ Yes votes in $V$. 
Since $y$ beats $w$ in the run-off, this contradicts the fact that
$w$ wins the election.  It follows that there are at most $k$ voters
corresponding to $S_i$'s in $V_2$.
Thus, there are exactly $k$ such voters, and
these voters correspond to an exact cover.~\end{proofs}

\section{Conclusions}
In this paper, we studied the computational resistance and
vulnerability of three voting systems---plurality, Condorcet, and
approval voting---to 
destructive control by an
election's chair in each of 
seven control scenarios:
candidate addition, suppression, partition, and
run-off partition, and voter addition, suppression, and partition.
We classified each case as
immune, vulnerable, or computationally resistant.  We also 
studied the analogous constructive control cases and fully
resolved those  
that were not considered by Bartholdi,
Tovey, and Trick~\cite{bar-tov-tri:j:control}.

We identified cases where a system immune to constructive control
still can be vulnerable to destructive control (e.g., Condorcet voting
for control by adding candidates), and vice versa (e.g., approval voting
for control by deleting candidates).  We saw
that, among the systems studied, none is globally superior to the
others.  Rather, when choosing a voting system, one's choice will
depend on the types of control against which protection is most
desired.  Finally, we saw that---in contrast to some 
comments in earlier papers---tie-breaking is a far from minor
issue:
For those control types that involve partitions of
the candidate or voter set, we studied two natural tie-handling rules,
and we found specific cases in which the complexity of the
corresponding control problem varies crucially depending on which
tie-handling rule is adopted.

\bigskip

\noindent
{\bf Acknowledgments:}
We thank Jeroen Snippe
for helpful comments, and we thank
Klaus Wagner and his group for hosting a visit during
which this work was done in part.

\bibliographystyle{alpha}

\appendix

\section{Appendix: Corrigendum to the
  Proof of
  Theorem 4.21 in Both the Journal Version and
  the Most Recent Previous TR Version%
\protect\footnote{Supported in part by
 Renewed Research Stay grants from the Alexander von Humboldt Foundation
 and by DFG grant RO~1202/21-2 (project number 438204498).
 This corrigendum was done in part while 
 the first two authors were visiting
 Heinrich-Heine-Universit\"at D\"usseldorf.}}

This corrigendum applies to both the most recent previous
technical report version of the paper
(arXiv:cs/0507027v4, March~2006)
and the journal version
(Artificial Intelligence, V.~171, \#5--6, pp.~255--285, April~2007).
That is because 
both have identical proofs of Theorem~4.21---which is numbered
Theorem~4.21 in
both of them---right down to having 
the same
example and equation numbers.
Thus the
second half of each is flawed in the same way, which will be 
corrected in this corrigendum.
This proof also
replaces the analogous
proof part of
Theorem~6 of the AAAI version.

The argument in the ``Loop in model TP'' section in the proof of
Theorem~4.21 of ``Anyone but Him: {The} Complexity of Precluding an
Alternative''---which is part of the paper's proof that approval
voting is vulnerable to destructive control by partition of voters in
the TP (ties promote) first-round model and the unique-winner
final-round model---does not correctly handle cases where one or both
of $S_a$ and $S_b$ are zero;\footnote{We refer to the journal
  version---or, essentially equivalently, Version~4 of the technical report---%
for all notation not defined here.
In particular, when clear from context, we will
use the vote counts (such as~$S_{ac}$) to refer not only to the counts
but also to the collection of that type of vote. 
}
in particular
the algorithm can give an incorrect
answer in such cases. 

For example, consider the case with these as the bins having positive numbers of votes:
$S_{ac}=2$, $S_{bc}=3$, $S_a=1$, and $L_c=1$.
In this 7-vote case, the ``Checking the trivial 
cases'' screening step in the paper does not 
step in (note that $c$ is the unique winner 
in $(C,V)$ and there are at least three candidates),
and the paper's 3-part disjunctive test to
identify  ``hopeless'' cases
claims that control is impossible in this case (i.e., the chair cannot 
prevent $c$ from being a unique winner), since 
$S_b = 0$.
Yet
control clearly is possible (i.e., the 
chair can ensure that $c$ is not a unique
winner) in this case, simply putting into
$V_1$ the two $S_{ac}$ votes and the sole vote in
$S_a$, and putting the other four 
votes into $V_2$.  (If one wants a 
counterexample in which 
\emph{both} $S_a$ and $S_b$ equal 
zero, the following 8-vote example 
provides one:
$S_{ac}=S_{bc}=3$ and $L_c=2$.)

The modified proof below changes
the 3-part disjunctive
expression used to screen off 
a set of cases where control is so obviously hopeless
that the loop need not be called
in those cases.
The changed/corrected 3-part disjunctive expression 
now no longer erroneously 
eliminates cases where control in fact
can succeed, though the corrected version unfortunately
uses a more complex and more
restrictive expression.

The fact that we have already eliminated
those cases that satisfy the 3-part disjunctive test is used
heavily 
within the argument below of the correctness of the algorithm being
given. In fact, we will show that for 
all cases that are not shot down by 
the 3-part disjunctive test, control
is possible, and we determine and output
a successful control action to show
\emph{certifiable} vulnerability as claimed).

We now give the revised construction and proof of correctness for the ``Loop in model TP'' section of the paper's proof of Theorem~4.21.

\noindent
{\bf Loop in model~TP:} For each $a, b \in C$ with
$||\{a,b,c\}||=3$, 
we test whether we can make $a$ strictly beat $c$ in
$\parpair{C,V_1}$ while also making $b$ strictly beat $c$ in $\parpair{C,V_2}$
for some partition of $V$ into $V_1$ and~$V_2$.  
(We do not need to consider the case where $a$ and $b$ are the
same, since if the same candidate strictly beats $c$ in both
subelections, then $c$ is certainly not a unique winner of $(C,V)$,
and so this case would have been handled already within the ``Checking
the trivial cases'' step that runs before this ``loop'' step.)

Throughout, we'll (as noted within the paper in its setup for this proof) ignore votes that approve of all
of $a$, $b$, and $c$, or that disapprove of all of them; those votes
can go in either side of the partition, since they have no effect on
how $a$ and $b$ do relative to $c$.

If $W_c - L_c > S_a + S_b - 2$ or $((S_a=0 \lor S_b=0)\land L_c=0)$,
then this $a$ and $b$ are hopeless, so move on
to consider the next $a$ and $b$ in the loop.

For all other cases, will show  that there is
a successful partition, i.e., a partition
that ensures that $c$ does not uniquely
win. So let us assume that above ``if'' fails,
which means we have that 
\begin{equation}
  W_c - L_c  \leq  S_a + S_b - 2 \label{A1}
\end{equation}
and
\begin{equation}
(S_a>0 \land S_b>0)\lor L_c>0.\label{A2}
\end{equation}

Of course, our best approach is to put all $S_{ac}$ and $S_a$ votes
into $V_1$ and to put all $S_{bc}$ and $S_b$ votes into~$V_2$, since if
there is any successful partition, there clearly is one that does
that. So we now will assume that those 
will be always assigned by the chair in
that way. Note that due to this, 
we will not in our discussions below
need to mention $S_{ac}$ or~$S_{bc}$,
because---given that they 
are put into $V_1$ and~$V_2$, respectively---they in those 
have no effect on the difference in 
votes between $a$ and $c$ in $V_1$ 
or between $b$ and $c$ in~$V_2$.

So ``all'' that remains is to see if there is a placement of the
$L_c$ and $W_c$ votes that ensures that
$a$ gets (strictly) more votes
than $c$ in $V_1$ and that 
$b$ gets (strictly) more votes
than $c$ in $V_2$.

Let us go through the cases.

If $S_a=S_b=0$, then we know by (\ref{A1}) that $L_c \geq 2$.
Assign one vote from $L_c$ to $V_1$ and the
rest to~$V_2$, and assign all the votes 
from $W_c$ to~$V_2$.
So
$a$ beats $c$ in $(C,V_1)$ by
$S_a + 1 = 1>0$ vote, and 
$b$ beats $c$ in $(C,V_2)$ by
\[
S_b + (L_c-1) - W_c \geq_{\text{by~(\ref{A1})}} 2-S_a - 1 = 1>0
\]
votes.

If exactly one of $S_a$ and $S_b$ equals zero, let us suppose that we
have $S_a=0$, and $S_b > 0$ (the case $S_a>0$, and $S_b =0$ is
analogous aside from naming, and so we do not need to discuss it separately).
By~(\ref{A2}), we must have $L_c>0$.
Assign one vote from $L_c$ to $V_1$ and the rest to $V_2$,
and assign all of $W_c$ to~$V_2$.
So $a$ beats $c$ in $(C,V_1)$ by
$S_a + 1 = 1>0$ vote, and
$b$ beats $c$ in $(C,V_2)$ by
\[
S_b + (L_c-1) - W_c \geq_{\text{by~(\ref{A1})}}
1 - S_a = 1-0 > 0
\]
votes.

Finally, consider the case $S_a>0$ and $S_b>0$.
Without loss of generality, assume $S_b \leq S_a$ (otherwise,
exchange the names of $a$ and $b$ so that that holds).
Put $\min(W_c,S_b-1)$ votes from $W_c$ into $V_2$
and put all of $L_c$ and also the remaining
$W_c - \min(W_c,S_b-1)$ members of $W_c$ into $V_1$.
So $b$ beats $c$ by at least
\[
S_b - \min(W_c,S_b-1) \geq S_b - (S_b-1) = 1 > 0
\]
approvals in $(C,V_2)$.
Also, $a$'s approvals minus $c$'s approvals in $(C,V_1)$ total
$S_a + L_c - (W_c - \min(W_c,S_b - 1))$.
If $\min(W_c,S_b - 1) = W_c$, then
\[
S_a + L_c - (W_c - \min(W_c,S_b - 1)) = S_a+L_c - 0
\]
which, since $S_a>0$ and $L_c \geq 0$, is greater than~$0$.
If $\min(W_c,S_b - 1) = S_b-1$ (the case where the values in
the ``min'' expression are equal to each other
is handled equally well by this
sentence and the previous one), then 
\[
S_a + L_c - (W_c - \min(W_c,S_b - 1)) = S_a+L_c - W_c+S_b - 1
\geq_{\text{by~(\ref{A1})}}
2-1 > 0.
\]

\begin{samepage}
This completes the subcases of analyzing the loop iteration
for the candidate pair $a$ and $b$.

\paragraph*{Acknowledgements for the Corrigendum}We thank Kerstin Neu for finding and bringing to our attention an example on which the original paper's
Theorem 4.21 ``Loop in model TP'' construction fails.

\end{samepage}
\end{document}